\documentclass[prd,%
preprint,%
tightenlines,superscriptaddress,showpacs,%
nofootinbib,eqsecnum,amsfonts,amsmath]{revtex4}
\usepackage{bm}
\usepackage{graphicx}

\begin{document}

\title{Surface-integral expressions for the multipole moments of
post-Newtonian sources and the boosted Schwarzschild solution}

\date{\today}

\author{Luc Blanchet}
\email{blanchet@iap.fr}\affiliation{${\mathcal{G}}{\mathbb{R}}
\varepsilon{\mathbb{C}}{\mathcal{O}}$, Institut d'Astrophysique de
Paris, C.N.R.S.,\\ 98$^{\text{bis}}$ boulevard Arago, 75014 Paris,
France}\affiliation{Yukawa Institute for Theoretical Physics, Kyoto
University, Kyoto 606-8502, Japan}

\author{Thibault Damour} \email{damour@ihes.fr} \affiliation{Institut
des Hautes \'Etudes Scientifiques, 35 route de Chartres, 91440
Bures-sur-Yvette, France}

\author{Bala R Iyer} \email{bri@rri.res.in} \affiliation{Raman
Research Institute, Bangalore 560 080, India}

\date{\today} 

\begin{abstract} 
New expressions for the multipole moments of an isolated
post-Newtonian source, in the form of surface integrals in the outer
near-zone, are derived. As an application we compute the ``source''
quadrupole moment of a Schwarzschild solution boosted to uniform
velocity, at the third post-Newtonian (3PN) order. We show that the
consideration of this boosted Schwarzschild solution (BSS) is enough
to uniquely determine one of the ambiguity parameters in the recent
computation of the gravitational wave generation by compact binaries
at 3PN order: $\zeta=-7/33$. We argue that this value is the only one
for which the Poincar\'e invariance of the 3PN wave generation
formalism is realized. As a check, we confirm the value of $\zeta$ by
a different method, based on the far-zone expansion of the BSS at
fixed retarded time, and a calculation of the relevant non-linear
multipole interactions in the external metric at the 3PN order.
\end{abstract}

\pacs{04.25.-g, 04.30.-w}

\maketitle

\section{Introduction}\label{intro}

The first aim of the present paper is to obtain, in the continuation
of previous work \cite{B95,B98mult}, some new expressions for the
multipole moments of an isolated gravitating source within the
post-Newtonian (PN) approximation scheme of general relativity. The
``\textit{source}'' multipole moments, as they were derived in
Ref.~\cite{B98mult}, are given, for any multipolar order $\ell$, by
certain functionals of the PN expansion of the pseudo-tensor of the
matter and gravitational fields, formally valid up to any PN
order. The moments have been obtained by solving the matching equation
between the inner PN field and the outer multipolar expansion. They
represent the generalization of the 1PN multipole moments obtained
earlier in Refs.~\cite{BD89,DI91b,DI91a}.\,\footnote{As usual the
$n$PN approximation refers to all the terms up to the relative order
$\sim 1/c^{2n}$, where $c$ is the speed of light. Powers of $c$ and
the gravitational constant $G$ will generally be explicitly displayed
here.} The source moments are the ones which parametrize the
``\textit{multipolar-post-Minkowskian}'' (MPM) expansion of the
external field of the source, as it was defined and investigated in
Ref.~\cite{BD86}.

In Ref.~\cite{B98mult} the PN moments have been given in the form of
some \textit{volume integrals}, whose support is non-compact because
of the contribution of the gravitational field, and extending
therefore over the full three-dimensional space. However, as we recall
below, their integrand is made from the PN expansion of the
stress-energy pseudo tensor $\tau^{\mu\nu}$, whose physical validity
is restricted to the \textit{near-zone} of the source. The formal
extension of the integrals to $r\equiv\vert\mathbf{x}\vert\rightarrow
+\infty$ was (uniquely) defined by introducing a suitable procedure of
analytic continuation.

In the present paper, we derive some equivalent expressions for the PN
source moments, in the form of \textit{surface integrals}, formally
performed at $r\rightarrow +\infty$. However let us emphasize that the
physical meaning of the limit $r \rightarrow +\infty$ (which is taken
after the PN limit $c \rightarrow +\infty$), corresponds to
considering what one can call the \textit{outer near-zone}, namely the
region which is at once far from the source, $r\gg$ (source radius),
and still within the near-zone, \textit{i.e.} $r\ll$ (wavelength). For
some physical problems, such alternative surface-integrals expressions
of the PN moments can be quite useful, as the next result of our paper
will illustrate.

Indeed, our second aim is to make a contribution to the problem of
theoretical templates of inspiralling compact binaries for
gravitational-wave experiments such as LIGO and VIRGO. Current
calculations of the gravitational waves are based on the expressions
of the PN source multipole moments \cite{B95,B98mult}, as well as on
high-order post-Minkowskian iteration of the external field
\cite{BD86,BD92,B98tail}. The mass-type quadrupole moment has been
computed in the case of compact binary systems up to the 2PN order
\cite{BDI95,WWi96} and more recently at 3PN order
\cite{BIJ02,BI04mult}. The physical motivation for such high-order PN
calculations of compact binary inspiral can be found in
Refs.~\cite{3mn,CF94,TNaka94,P95,DIS98,BCV03a,DIJS03}.

It has been shown in \cite{BIJ02} that at 3PN order the radiation
field of compact binaries, modelled as systems of point particles,
contains \textit{three ambiguity parameters}, $\xi$, $\kappa$ and
$\zeta$, due to an incompleteness of the Hadamard self-field
regularization, used in the computation of the quadrupole moment of
point particles at the 3PN order. In the present paper we shall show
how one can determine the value of one of these parameters, $\zeta$,
without any use of a self-field regularization scheme. The idea is to
consider the situation where one of the two masses is exactly
\textit{zero}. Such a limiting case corresponds to a single particle
moving with some uniform velocity. As it turns out, the ambiguity
parameter $\zeta$, but only this one, survives in this limit. We can
therefore compute it from the particular case of the 3PN quadrupole
moment generated by a single object, specifically a spherically
symmetric extended matter distribution, moving with uniform velocity
in the preferred reference frame with respect to which the multipole
moments are defined. The \textit{external} gravitational field of such
an object is evidently physically equivalent to a \textit{boosted
Schwarzschild solution} (BSS), \textit{i.e.} a Schwarzschild solution
viewed in a frame which is obtained from the usual ``Schwarzschild
rest frame'' by a Lorentz boost.

To compute the multipole moments of the BSS we propose and implement
two methods. The first one is to insert the metric of the BSS into our
new expressions for the PN multipole moments in terms of surface
integrals in the outer near-zone. The second method, substantially
more involved, consists of computing the far-zone or
``\textit{radiative}'' moments\,\footnote{Although the BSS does not
radiate gravitational waves, it is sometimes convenient to employ a
language for the different types of multipole moments which is similar
to the one used in the more general case of non-stationary matter
systems.} of the BSS by expanding the metric in the far zone at fixed
retarded time, and comparing the result with an analytic calculation
of the various non-linear multipole interactions occurring in the
far-zone moments up to the 3PN order. The source-type moment will
thereby be determined in an indirect way. We find that the results of
the two methods agree, and uniquely determine
\begin{equation}\label{zeta}
\zeta=-\frac{7}{33}\,.
\end{equation}

We argue below that (\ref{zeta}) is the \textit{unique} value for
which the 3PN wave generation formalism, as applied to binary systems
in Ref.~\cite{BIJ02,BI04mult}, incorporates the global
Lorentz-Poincar\'e invariance of general relativity. We have already
reported elsewhere \cite{BDEI04} that dimensional regularization leads
to the same value for $\zeta$ (as well as the determination of the
values of the two other ambiguity parameters $\xi$ and $\kappa$), and
therefore has the important feature of ``automatically'' preserving
the Poincar\'e invariance of the formalism. The calculation of $\zeta$
is similar to the one of its analogue in the 3PN equations of motion
of point particle binaries, namely the so-called ``kinetic'' ambiguity
constant $\omega_k$ \cite{JaraS98,JaraS99}, which has also been fixed
from the requirement of Poincar\'e invariance, either by using an
appropriate Lorentz-invariant version of Hadamard's regularization at
the level of the equations of motion \cite{BF00,BFeom}, or by direct
imposition that the ADM Hamiltonian be compatible with the existence
of phase-space generators satisfying the Poincar\'e algebra
\cite{DJSpoinc}. [The second and last ambiguity in the 3PN equations
of motion is the ``static'' ambiguity constant $\omega_s$
\cite{JaraS98,JaraS99} equivalent to the ambiguity parameter $\lambda$
\cite{BF00,BFeom}. Both $\omega_s$ and $\lambda$ (as well as
$\omega_k$ in fact) have been obtained by means of dimensional
regularization \cite{DJSdim,BDE04}. The same results have also been
achieved in Refs.~\cite{itoh1,itoh2} by expressing the equations of
motion at 3PN order in terms of surface integrals surrounding the
compact objects.]

The paper is organized as follows. Section \ref{mult} is devoted to a
general investigation of the PN source multipole moments of extended
matter sources. In Section \ref{recall} we present some useful
reminders of the formulation of the PN moments, and in Section
\ref{surfinf} we obtain our new expressions of these moments in terms
of surface integrals in the outer (or far) near-zone. Section
\ref{schw} is devoted to the investigation of the multipole moments of
the BSS. In Section \ref{sourcemoment} we apply the new expressions of
the PN source quadrupole moment to the case of the BSS and obtain the
crucial coefficient which yields Eq.~(\ref{zeta}) by comparison with
the BSS limit of the quadrupole moment of compact binaries. In Section
\ref{radiativemoment} we expand the field of the BSS at retarded
infinity and obtain the radiative-type quadrupole moment, which we
then relate to the source-type moment by computing the non-linear
multipole interactions therein up to 3PN order (the technical details
of the non-linear iteration are relegated to Appendix
\ref{nonlinear}). In this way we are able to confirm the value
(\ref{zeta}).

\section{Multipole moments as surface integrals}\label{mult}

\subsection{Reminders of the PN multipole moments}\label{recall}

As gravitational field variable we employ the standard ``Gothic metric
deviation'' from flat space-time, that we shall subject all over this
paper to the condition of harmonic coordinates, meaning that
\begin{subequations}\label{harmonic}\begin{eqnarray}
h^{\mu\nu}&\equiv& \sqrt{-g}\,g^{\mu\nu}-\eta^{\mu\nu}\,, \\
\partial_\nu h^{\mu\nu} &=& 0\,,
\end{eqnarray}\end{subequations}
where $g$ denotes the determinant and $g^{\mu\nu}$ the inverse of the
covariant metric: $g=\mathrm{det}(g_{\rho\sigma})$ and
$g^{\mu\nu}g_{\nu\sigma}=\delta^\mu_\sigma$, and where $\eta^{\mu\nu}$
is the Minkowski metric written in Cartesian coordinates:
$\eta^{\mu\nu}=\mathrm{diag}(-1,1,1,1)$. The Einstein field equations,
relaxed by the harmonic-coordinates condition, read as
\begin{equation}\label{EE}
\Box \,h^{\mu\nu} = \frac{16\pi G}{c^4} \,\tau^{\mu\nu}\,,
\end{equation}
where $\Box\equiv\Box_\eta$ is the \textit{flat space-time}
d'Alembertian or wave operator, and $\tau^{\mu\nu}$ denotes the total
stress-energy pseudo tensor of the matter and gravitational fields in
harmonic coordinates, given by
\begin{equation}\label{tau}
\tau^{\mu\nu} \equiv |g| \,T^{\mu\nu} + \frac{c^4}{16\pi G}
\,\Lambda^{\mu\nu}[h]\,.
\end{equation} 
Here $T^{\mu\nu}$ is the matter stress-energy tensor, and
$\Lambda^{\mu\nu}$ is proportional to the stress-energy distribution
of the gravitational field --- $\Lambda^{\mu\nu}$ is a functional of
the field strength $h$ or, more precisely, a function of $h$ and its
first and second space-time derivatives ($\sim\partial h$ and
$\partial^2h$), at least quadratic in $h$, $\partial h$ or
$\partial^2h$. The pseudo tensor $\tau^{\mu\nu}$ is conserved in a
Minkowskian sense,
\begin{equation}\label{dtau}
\partial_\nu\tau^{\mu\nu}=0\,,
\end{equation}
which is equivalent to the equation of motion of the matter source:
$\nabla_\nu T^{\mu\nu}=0$.

In principle one can solve Eq.~(\ref{EE}) by successive iterations in
the order of non-linearity, \textit{i.e.} by a formal expansion in
powers of Newton's constant $G$. This would constitute what one calls
the \textit{post-Minkowskian} expansion (with explicit consideration
of the matter source terms). However, such a straightforward
post-Minkowskian expansion does not lead to easily implementable
iterations (see, \textit{e.g.}, \cite{ThK75,CTh77}). Another scheme,
technically more useful, consists of splitting the problem of solving
(\ref{EE}) into three sub-problems. First, one solves the
\textit{vacuum} equations, obtained by setting the matter
stress-energy tensor $T^{\mu\nu}$ to zero in Eq.~(\ref{tau}), by a
particular post-Minkowskian expansion. Second, one solves the full
inhomogeneous Einstein equations (\ref{EE}) by a formal PN expansion,
in powers of the velocity of light $c$ (with $G$ fixed). And third,
one combines together these two expansions by means of an appropriate
variant of the method of matched asymptotic expansions.

The particular post-Minkowskian scheme we use to solve the vacuum
Einstein equations is the so-called
``\textit{multipolar-post-Minkowskian}'' (MPM) expansion of the metric
exterior to the source \cite{BD86}. This expansion combines a
non-linearity expansion in powers of $G$, with a multipolar expansion
of the successive non-linear iterations of $h^{\mu\nu}$, say
$h^{\mu\nu}_n$, \textit{i.e.}  essentially a decomposition of each
function $h^{\mu\nu}_n(t,r,\theta,\varphi)$ in tensorial spherical
harmonics on the unit sphere parametrized by $\theta$ and
$\varphi$. It will be convenient, following \cite{B98mult}, to denote
the MPM expansion of some quantity like $h^{\mu\nu}$, as
$\mathcal{M}(h^{\mu\nu})$, where the calligraphic letter $\mathcal{M}$
serves mainly the purpose of reminding the \textit{multipolar} aspect
of the expansion. [The post-Minkowskian aspect, though playing a
crucial role in the definition and especially the construction of MPM
metrics in \cite{BD86}, is less important to keep in mind in the
reasoning that we shall follow below.] The MPM expansion of the metric
is thus written as
\begin{equation}\label{hMPM}
\mathcal{M}(h^{\mu\nu}) = \sum_{n=1}^{+\infty} G^n\,h^{\mu\nu}_n\, ,
\end{equation}
where one must always have in mind that each term
$h^{\mu\nu}_n(t,r,\theta,\varphi)$ is decomposed in tensorial
spherical harmonics (for instance STF tensorial harmonics like in
\cite{BD86}). The successive post-Minkowskian coefficients
$h^{\mu\nu}_n$ are constructed iteratively from the linearized
approximation $h^{\mu\nu}_1$ by solving the harmonic-coordinates
vacuum field equations. These consist of the harmonicity condition
$\partial_\nu\mathcal{M}(h^{\mu\nu})=0$ and
\begin{equation}\label{boxhMPM}
\Box\,\mathcal{M}(h^{\mu\nu}) = \mathcal{M}(\Lambda^{\mu\nu})\,,
\end{equation}
where $\mathcal{M}(\Lambda^{\mu\nu})$ means $\Lambda^{\mu\nu}$ in
which we substitute for $h$, $\partial h$, $\partial^2h$ their
multipolar expansion: $\mathcal{M}(\Lambda^{\mu\nu})\equiv
\Lambda^{\mu\nu}[\mathcal{M}(h)]$. Again we stress that
$\mathcal{M}$(some field) refers to a quantity which is supposed to be
decomposed in a spherical-harmonics expansion (with coefficients being
some functions of $t$ and $r$).

The algorithm to explicitly construct the successive post-Minkowskian
coefficients $h^{\mu\nu}_n$, from the knowledge of the linearized
approximation $h^{\mu\nu}_1$, was given in Ref.~\cite{BD86}. As
everything depends on $h^{\mu\nu}_1$, the parametrization of the
linearized approximation will determine the full MPM metric
(\ref{hMPM}). In our approach the \textit{source} multipole moments
are defined as the ``seed moments'' that one introduces at the start
of the MPM scheme to parametrize the linearized approximation
$h^{\mu\nu}_1$.

With full generality one can write the linearized metric
$h^{\mu\nu}_1$, which satisfies $\Box\, h^{\mu\nu}_1=0$ together with
$\partial_\nu h^{\mu\nu}_1=0$, in the form
\begin{equation}\label{h1MPM}
h^{\mu\nu}_1=h^{\mu\nu}_{\mathrm{can}\,1}+\partial^\mu\varphi_1^\nu
+\partial^\nu\varphi_1^\mu-\eta^{\mu\nu}\partial_\lambda
\varphi_1^\lambda\,,
\end{equation}
where $\varphi_1^\mu$ is a linearized gauge transformation vector
(satisfying $\Box\,\varphi_1^\mu=0$, so the harmonic gauge condition
is preserved), and where $h^{\mu\nu}_{\mathrm{can}\,1}$ represents a
useful form of the linearized multipolar metric, called ``canonical''
and introduced in Ref.~\cite{Th80}. The mass-type and current-type
source multipole moments, $I_L(t)$ and $J_L(t)$, respectively, are
symmetric and trace-free (STF) tensors with respect to their $\ell$
indices ($\ell$ is the multipolar order). They parametrize the
canonical metric \textit{via} the following definition,
{\allowdisplaybreaks \begin{subequations}\label{h1can}\begin{eqnarray}
h^{00}_{\mathrm{can}\,1} &=& -\frac{4}{c^2}\sum_{\ell = 0}^{+\infty}
\frac{(-)^\ell}{\ell !}  \partial_L \left[ \frac{1}{r} I_L
(u)\right]\,,\\ h^{0i}_{\mathrm{can}\,1} &=& \frac{4}{c^3}\sum_{\ell =
1}^{+\infty} \frac{(-)^\ell}{\ell !}  \left\{ \partial_{L-1} \left[
\frac{1}{r} \dot{I}_{iL-1} (u)\right] + \frac{\ell}{\ell+1}
\varepsilon_{iab} \,\partial_{aL-1} \left[ \frac{1}{r} J_{bL-1}
(u)\right]\right\}\,,\\ h^{ij}_{\mathrm{can}\,1}
&=&-\frac{4}{c^4}\sum_{\ell = 2}^{+\infty} \frac{(-)^\ell}{\ell !}
\left\{ \partial_{L-2} \left[ \frac{1}{r}\ddot{I}_{ijL-2} (u)\right] +
\frac{2\ell}{\ell+1} \partial_{aL-2} \left[ \frac{1}{r}
\varepsilon_{ab(i} \dot{J}_{j)bL-2} (u)\right]\right\}\,.
\end{eqnarray}
\end{subequations}}\noindent
Our notation is standard.\,\footnote{$L\equiv i_1\cdots i_\ell$
denotes a multi-index composed of $\ell$ multipolar indices $i_1,
\cdots, i_\ell$; $\partial_L\equiv \partial_{i_1}\cdots
\partial_{i_\ell}$ means a product of $\ell$ partial derivatives
$\partial_i\equiv \partial/\partial x^i$; similarly $x_L\equiv
x_{i_1}\cdots x_{i_\ell}$ is a product of $\ell$ spatial vectors
$x_i\equiv x^i$; symmetric-trace-free products are denoted with hats
so that $\hat{x}_L\equiv \mathrm{STF}(x_L)$; sometimes we shall also
use some brackets surrounding the STF indices: $x_{\langle
L\rangle}\equiv\hat{x}_L$; the dots refer to the partial time
derivation; $\varepsilon_{abi}$ is the Levi-Civita totally
antisymmetric symbol (such that $\varepsilon_{123}=1$); index
symmetrization means $(ij)\equiv \frac{ij+ji}{2}$;
$r\equiv\vert\mathbf{x}\vert$ and $u\equiv t-r/c$.} In addition, the
components of the gauge transformation vector $\varphi_1^\mu$ can be
parametrized by four other sequences of multipole moments, called
$W_L$, $X_L$, $Y_L$ and $Z_L$, also being STF in their indices $L$, in
the way specified by Eq.~(4.13) in \cite{B98mult}, \textit{i.e.}
{\allowdisplaybreaks
\begin{subequations}\label{phi1}
\begin{eqnarray}
\varphi^0_1 &=& \frac{4}{c^3}\sum_{\ell = 0}^{+\infty}
\frac{(-)^\ell}{\ell !}  \partial_L \left[ \frac{1}{r} W_L
(u)\right]\,,\\ \varphi^i_1 &=& -\frac{4}{c^4}\sum_{\ell =
0}^{+\infty} \frac{(-)^\ell}{\ell !}  \partial_{iL} \left[ \frac{1}{r}
X_L (u)\right] \nonumber \\ &&-\frac{4}{c^4}\sum_{\ell = 1}^{+\infty}
\frac{(-)^\ell}{\ell !} \left\{ \partial_{L-1} \left[ \frac{1}{r}
Y_{iL-1} (u)\right] + \frac{\ell}{\ell+1} \varepsilon_{iab}
\partial_{aL-1} \left[ \frac{1}{r} Z_{bL-1} (u)\right]\right\}\,.
\end{eqnarray}
\end{subequations}}\noindent

The complete set of moments parametrizing the linearized approximation
(\ref{h1MPM}) with (\ref{h1can}) and (\ref{phi1}), \textit{i.e.}
$\{I_L,\,J_L,\,W_L,\,X_L,\,Y_L,\,Z_L\}$, is collectively referred to
as the \textit{source} multipole moments. All these moments are
defined in such a way that they admit a non-zero finite Newtonian
limit, when $c\rightarrow +\infty$. It is clear that the most
important of these moments are the mass-type moment $I_L$ and the
current-type one $J_L$. Indeed the other moments,
$W_L,\,\cdots,\,Z_L$, parametrize a gauge transformation and thus do
not play any physical role at the linearized order (though they do
play a role at the non-linear level). 

In Section \ref{radiativemoment} we shall also recall the definition
of two and only two sets of moments, named the ``canonical'' moments,
denoted by $\{M_L,\,S_L\}$, which are physically equivalent to the
complete set of \textit{six} source moments
$\{I_L,\,J_L,\,\cdots,\,Z_L\}$, in the sense that they describe the
\textit{same} external gravitational field. However, following
\cite{B98mult}, we prefer to reserve the name of source moments to the
set $I_L,\,J_L,\,\cdots,\,Z_L$ because they are connected \textit{via}
some analytic closed form expressions to the stress-energy tensor of a
PN source.

The MPM algorithm computes sequentially any of the non-linear
coefficients in Eq.~(\ref{hMPM}) as follows
\cite{BD86,B98quad}.\,\footnote{In this paper we adopt a slightly
modified version of the MPM algorithm, defined in Section 2 of
Ref.~\cite{B98quad}, which is more convenient in practical
computations.}  Suppose that the first $n-1$ coefficients
$h_1,\,\cdots,\,h_{n-1}$, where $h_1$ is given by
Eqs.~(\ref{h1MPM})-(\ref{phi1}), have been constructed. We have then
to solve, at the $n$-th order, the inhomogeneous wave equation,
\begin{equation}\label{boxhn}
\Box\, h^{\mu\nu}_n = \Lambda^{\mu\nu}_n\,,
\end{equation}
whose source term is known from the previous iterations, \textit{i.e.}
$\Lambda^{\mu\nu}_n=\Lambda^{\mu\nu}_n[h_1,\,\cdots,\,h_{n-1}]$, and
where as always we also have to satisfy the coordinate condition
$\partial_\nu h^{\mu\nu}_n=0$. The solution, satisfying a condition of
``stationarity in the past'' ensuring that the correct boundary
condition at Minkowskian past null-infinity are satisfied, reads
\begin{equation}\label{hn}
h^{\mu\nu}_n = u^{\mu\nu}_n + v^{\mu\nu}_n\,.
\end{equation}
The first term represents the standard retarded integral operator
(denoted $\Box^{-1}_\mathrm{R}$ below) acting on the non-linear
source, but augmented by a specific regularization scheme to deal with
the divergencies of the retarded integral introduced by the fact that
the multipolar expansion diverges at the origin of the spatial
coordinates, when $r\equiv\vert\mathbf{x}\vert\rightarrow 0$ [see
\textit{e.g.} Eq.~(\ref{h1can})]. Posing
\begin{equation}\label{un}
u^{\mu\nu}_n =
\mathop{\mathrm{FP}}_{B=0}\,\Box^{-1}_\mathrm{R}\left[\Bigl(
\frac{r}{r_0}\Bigr)^B\,\Lambda^{\mu\nu}_n\right]\,,
\end{equation}
we do solve the required wave equation, \textit{i.e.} $\Box\,
u^{\mu\nu}_n = \Lambda^{\mu\nu}_n$, provided that the finite part (FP)
takes the meaning specified below. The second term in Eq.~(\ref{hn})
represents a particular homogeneous solution, \textit{i.e.} $\Box\,
v^{\mu\nu}_n = 0$, defined in such a way that the harmonic gauge
condition $\partial_\nu h^{\mu\nu}_n = 0$ is satisfied at this order
(see Refs.~\cite{BD86,B98quad} for the details).

To define the FP process we multiply the source term in Eq.~(\ref{un})
by a factor $(r/r_0)^B$, where $B\in\mathbb{C}$ and $r_0$ denotes some
arbitrary length scale, we compute the $B$-dependent retarded integral
in the domain of the complex $B$ plane in which it converges,
\textit{i.e.}  for which $\Re(B)$ is initially a large enough positive
number, and we define it in a neighborhood of the value of interest
$B=0$ by analytic continuation. The finite part in (\ref{un}) means
the coefficient $a_0$ of the zero-th power of $B$ in the Laurent
expansion $\sum a_p \,B^p$ (where $p\in\mathbb{Z}$) of the retarded
integral when $B\rightarrow 0$. We emphasize that the divergencies
cured by the FP regularization in (\ref{un}) are ultraviolet (U.V.)
type divergencies ($r\rightarrow 0$). As we shall see the same FP
regularization will be used in the multipole moments to deal with
their infra-red (I.R.)  divergencies (when $r\rightarrow +\infty$).

In order to prevent any confusion, let us clarify the meaning of the
various expansions that we shall use, and of the limits $r\rightarrow
0$ and $r\rightarrow +\infty$ that will arise in the present
paper. First, we recall that any MPM-expanded quantity is always
thought of as written as a double expansion: one expansion in powers
of $G$ and one in spherical harmonics, say $h_n=\sum_\ell \hat{n}_L
F_L$, where $L$ denotes a multi-index which carries an irreducible
representation of the rotation group (we do not write space-time
indices). Then, each of the coefficients of this double expansion is
given by some explicit function of $t$ and $r$, say $F_L(t,r,c)$,
where we have also indicated a dependence on the velocity of light $c$
(\textit{cf.} for instance the simple case of the linearized
approximation above). An important technical aspect of our formalism
is that we shall consider these functions $F_L(t,r,c)$ in the
\textit{whole range} of the radial coordinate $r$, even if this range
does not correspond to a space-time region where the corresponding
expansion is physically valid. Moreover, we shall sometimes consider
instead of the original MPM coefficients $F_L(t,r,c)$ their PN
expansion (or ``near-zone expansion''), which means technically a
formal expansion in powers of $1/c$ (with possibly some powers of $\ln
c$). We shall denote the PN expansion of any quantity by an
overline. For instance, $\overline{F}_L(t,r,c)$ denotes the expansion
in powers of $1/c$ of $F_L(t,r,c)$, when keeping fixed the variables
$r$ and $t$.
 
We deal sometimes [as in Eq.~(\ref{un}) above] with functions
$F_L(t,r,c)$ in the region $r\rightarrow 0$, which is mathematically
well defined (by real analytic continuation in $r$) but which
physically corresponds to a region where the \textit{vacuum} MPM
metric should be replaced by a solution of the \textit{inhomogeneous}
Einstein equations, and therefore where the actual physical function
would be a different function of $r$ (and $t$) than $F_L(t,r,c)$. On
the other hand, the limit $r\rightarrow +\infty $ is acceptable both
mathematically and physically for $F_L(t,r,c)$, because it comes from
a post-Minkowskian expansion which is valid all over the exterior of
the source. However, we shall also mathematically deal with the PN
re-expansion $\overline{F}_L(t,r,c)$ of the function $F_L(t,r,c)$, and
then formally consider the function $\overline{F}_L(t,r,c)$ in the
limit $r\rightarrow +\infty$. Mathematically, the behavior of
$\overline{F}_L(t,r,c)$ when $r\rightarrow +\infty$ is again well
defined (by real analytic continuation in $r$) for each term
$F_L(t,r,c)$ in the MPM expansion. Though the limit $r\rightarrow
+\infty$ seems physically incorrect for a PN expansion, it is here
technically (or mathematically) well-defined. We stress that the PN
limit $c \rightarrow +\infty$ is taken \textit{before} considering the
$r\rightarrow +\infty$ behavior. If we remember that the PN expansion
is physically valid only in the \textit{ near-zone} of the source,
defined as $ r\ll c\,T$, where $T$ is a characteristic time of
variation of the source, we see that the physical domain of validity
of the $r\rightarrow +\infty$ expansion of the PN expanded functions
$\overline{F}_L(t,r,c)$ actually corresponds to the \textit{outer}
part of the near-zone, \textit{i.e.} when $r$ is much larger than the
size, say $a$, of the source (multipole expansion), but still
significantly smaller than a gravitational wave-length $\lambda\sim
c\,T$. We shall often refer to this domain as the \textit{far
near-zone}.

As explained in Ref.~\cite{BD86}, in order to be able to define the
behavior for $r\rightarrow 0$ and $r\rightarrow +\infty$ of all the
coefficients $F_L(t,r,c)$ appearing in the successive iterations of
the MPM scheme one needs to make some formal technical assumptions:
One must start with only a finite number of ``seed'' multipole
moments, assume that they are infinitely differentiable functions of
time, and that they tend to some constants in the infinite past. In
our approach we assume that these requirements are initially
satisfied, and we formally take, at the end of the calculation, a
limit where these requirements are relaxed (so that we extend our
results to an infinite number of moments, which are not necessarily
past-stationary).

The MPM expansion, sketched above, must be completed by an expansion
scheme which covers the source. This is done by considering also a PN
approximation for solving the inhomogeneous Einstein equations. The PN
scheme is \textit{a priori} valid in the near-zone ($r\ll c\,T$),
while the MPM one is valid in the exterior of the source ($r >
a$). The two domains of validity overlap in the exterior near-zone ($a
< r\ll c\,T$). One then imposes a ``matching condition'' which will
enable us to determine the values of the multipole moments as
functionals of the PN source. If we denote as above by
$\overline{h}^{\mu\nu}$ the PN expansion the matching condition can be
expressed as
\begin{equation}\label{match}
\mathcal{M}\bigl(\overline{h}^{\mu\nu}\bigr)\equiv
\overline{\mathcal{M}\left(h^{\mu\nu}\right)}\,,
\end{equation}
which says that the multipolar re-expansion of the PN metric
$\overline{h}^{\mu\nu}$ agrees, in the sense of formal series, with
the \textit{near-zone} re-expansion (also denoted with an overbar) of
the MPM metric $\mathcal{M}\left(h^{\mu\nu}\right)$. If we consider
that a multipolar expansion is essentially an expansion in inverse
powers of $r$ when $r$ gets far-away from the source, we can roughly
summarize the matching equation (\ref{match}) as saying that the far
expansion ($r\rightarrow +\infty,\,t=\mathrm{const}$) of the near-zone
metric --- L.H.S. of Eq.~(\ref{match}) --- coincides with the near
expansion ($r/ c\,T \rightarrow 0,\,t=\mathrm{const}$) of the far
(multipolar-expanded) metric --- R.H.S. of (\ref{match}). The common
general structure of both sides of (\ref{match}) will be given in
Eqs.~(\ref{Mhexp}) and (\ref{MhexpNZ}).

In Ref.~\cite{B98mult} the PN source moments were obtained as
functionals of the PN expansion of the pseudo-stress energy tensor
defined by Eq.~(\ref{tau}), namely $\overline{\tau}^{\mu\nu}$. For the
main source moments $I_L$ and $J_L$ (with any $\ell\geq 2$) we get
{\allowdisplaybreaks
\begin{subequations}\label{ILJL}\begin{eqnarray}\label{IL} I_L(t)&=&
\frac{1}{c^2}\,\mathop{\mathrm{FP}}_{B=0}\int
d^3\mathbf{x}\,\Bigl(\frac{r}{r_0}\Bigr)^B \int^1_{-1} dz\left\{
\delta_\ell(z) \,\hat{x}_L\,\left(\overline{\tau}^{00}
+\overline{\tau}^{ii}\right)(\mathbf{x},t+z\,r/c)\right.  \nonumber\\
&&\qquad\qquad \left. -\frac{4(2\ell+1)}{(\ell+1)(2\ell+3)}
\,\delta_{\ell+1}(z) \,\hat
x_{iL}\,\frac{\partial\overline{\tau}^{i0}}{c\partial
t}(\mathbf{x},t+z\,r/c)\right.  \nonumber\\ &&\qquad\qquad
\left. +\frac{2(2\ell+1)}{(\ell+1)(\ell+2)(2\ell+5)}
\,\delta_{\ell+2}(z) \,\hat{x}_{ijL}
\,\frac{\partial^2\overline{\tau}^{ij}}{c^2\partial
t^2}(\mathbf{x},t+z\,r/c) \right\}\,,\\\label{JL} J_L(t)&=&
\frac{1}{c}\,\mathop{\mathrm{FP}}_{B=0}\varepsilon_{ab\langle i_\ell}
\int d^3\mathbf{x}\,\Bigl(\frac{r}{r_0}\Bigr)^B \int^1_{-1} dz\left\{
\delta_\ell(z)\,\hat{x}_{L-1\rangle a}
\overline{\tau}^{b0}(\mathbf{x},t+z\,r/c)\right.  \nonumber\\
&&\qquad\qquad \left.
-\frac{2\ell+1}{(\ell+2)(2\ell+3)}\,\delta_{\ell+1}(z)
\,\hat{x}_{L-1\rangle ac}
\frac{\partial\overline{\tau}^{bc}}{c\partial
t}(\mathbf{x},t+z\,r/c)\right\}\,,
\end{eqnarray}\end{subequations}}\noindent
where we recall that $\hat{x}_L$ means the STF product of $\ell$
spatial vectors, $\hat{x}_L\equiv \mathrm{STF}(x_{i_1}\cdots
x_{i_\ell})$. The other source moments, $W_L,\,X_L,\,Y_L,\,Z_L$, are
given by Eqs.~(5.17)-(5.20) in Ref.~\cite{B98mult}. We shall give
below their new expressions in terms of surface integrals. See also
\cite{B98mult} for a discussion of the conserved monopole and dipole
moments (having $\ell=0,1$).

A basic feature of these expressions is that the integral formally
extends over the whole support of the PN expansion of the
stress-energy pseudo-tensor, \textit{ i.e.}  from $r=0$ up to
infinity. As already emphasized, the formal series
$\overline{\tau}^{\mu\nu}$ is physically meaningful only within the
near-zone. Therefore the integrals (\ref{ILJL}) physically refer to a
result obtained from near-zone quantities only (in the formal limit
where $c \rightarrow +\infty$). However, it was found convenient in
Ref.~\cite{B98mult} to mathematically extend the integrals up to
$r\rightarrow +\infty$. This was made possible by the use of the
prefactor $(r/r_0)^B$, together with a process of analytic
continuation in the complex $B$ plane. This shows up in
Eqs.~(\ref{ILJL}) as the crucial Finite Part (FP) operation, when
$B\rightarrow 0$, which technically allows one to uniquely define
integrals which would otherwise be I.R. divergent, \textit{i.e.}
divergent at their upper boundary, $\vert\mathbf{x}\vert\rightarrow
+\infty$. See Refs.~\cite{B95,B98mult} for the proof and details.

Since Eqs.~(\ref{ILJL}) are valid only in the sense of PN expansions,
the operational meaning of the auxiliary integrals in (\ref{ILJL}),
with respect to the variable $z$, is actually that of an infinite PN
series, given by
\begin{subequations}\label{intdeltal}\begin{eqnarray}
\int^1_{-1} dz\,\delta_\ell(z) \,\overline{\tau}^{\mu\nu}(\mathbf{x},
t+z\,r/c) &=& \sum_{k=0}^{+\infty}\,\alpha_{k,\ell}
\,\left(\frac{r}{c}\frac{\partial}{\partial t}\right)^{2k}
\overline{\tau}^{\mu\nu}(\mathbf{x},t)\,,\\
\hbox{where}\quad\alpha_{k,\ell} &\equiv&
\frac{(2\ell+1)!!}{(2k)!!(2\ell+2k+1)!!}\,.\label{alphakl}
\end{eqnarray}\end{subequations}
The expression of the function $\delta_\ell (z)$, given in Appendix B
of Ref.~\cite{BD89} as
\begin{equation}\label{deltal}
\delta_\ell (z) \equiv \frac{(2\ell+1)!!}{2^{\ell+1} \ell!}
(1-z^2)^\ell~~\hbox{such that}\int_{-1}^{1} dz\,\delta_\ell(z) = 1\,,
\end{equation}
is useful when manipulating formal PN expansions such as
(\ref{intdeltal}), but will not be used explicitly in the present
investigation.

\subsection{The multipole moments as surface integrals}\label{surfinf}
 
In this Section we derive an alternative form of the PN source moments
(\ref{ILJL}) in terms of two-dimensional surface integrals. Such a
possibility of expressing the moments, for general $\ell$ and at any
PN order, as some surface integrals is quite useful for practical
purposes, as we shall show below when considering the application to
the BSS case. In keeping with the fact, just explained, that the
``volume integrals'' Eqs.~(\ref{ILJL}) physically involve only
near-zone quantities, the ``surface integrals'' into which we shall
transform Eqs.~(\ref{ILJL}) physically refer to an operation which
extracts some coefficients in the ``far near-zone'' expansion of the
gravitational field, \textit{i.e.} in the expansion in increasing
powers of $1/r$ of the PN-expanded near-zone metric. Technically, as
our starting point (\ref{ILJL}) is made of integrals extended up to
$r\rightarrow +\infty$, our mathematical manipulations below will
involve ``surface terms'' on arbitrary large spheres $r =
\mathcal{R}$. All our manipulations will be mathematically
well-defined because of the properties of complex analytic
continuation in $B$.

The basic idea is to go from the ``source term'',
$\overline{\tau}^{\mu\nu}$, to the corresponding ``solution''
$\overline{h}^{\mu\nu}$, \textit{via} integrating by parts the Laplace
operator present in $\overline{\tau}^{\mu\nu}=\frac{c^4}{16\pi
G}\,\Box\,\overline{h}^{\mu\nu}$. From Eq.~(\ref{intdeltal}) we have
\begin{equation}\label{intdeltal0}
\int d^3\mathbf{x}\,r^B \,\hat{x}_L\,\int^1_{-1} dz~ \delta_\ell(z)
\,\overline{\tau}^{\mu\nu} = \frac{c^4}{16\pi
G}\sum_{k=0}^{+\infty}\,\alpha_{k,\ell} \,\left(\frac{d}{c
dt}\right)^{\!2k}\int
d^3\mathbf{x}\,r^{B+2k}\,\hat{x}_L\,\Box\,\overline{h}^{\mu\nu}\,,
\end{equation}
in which we insert $\Box=\Delta-\left(\frac{\partial}{c\,\partial
t}\right)^2$ on the R.H.S., and operate the Laplacian by parts using
$\Delta(r^{B+2k}\,\hat{x}_L)=(B+2k)(B+2\ell+2k+1)r^{B+2k-2}\,\hat{x}_L$. In
the process we can ignore the all-integrated surface terms because
they are identically zero by complex analytic continuation, from the
case where the real part of $B$ is chosen to be a large enough
\textit{negative} number. [The complete justification of this is as
follows. In the present formalism we are actually dealing with the MPM
metric given by the non-linearity expansion (\ref{hMPM}), and we are
working with the calculation of some given \textit{finite}
post-Minkowskian approximation $n$. Then the PN expansion of the
post-Minkowskian metric coefficient $h_n^{\mu\nu}$, namely
$\overline{h}_n^{\mu\nu}$, will typically diverge at infinity, but not
more than a certain finite power of $r$, say $N(n)$, depending on $n$
and such that $\lim_{n\rightarrow \infty}N(n)=\infty$. Using
$\overline{h}_n^{\mu\nu}=\mathcal{O}\left(r^{N(n)}\right)$ it is then
clear that the all-integrated terms in question are zero when we
choose initially $\Re (B)+2k+\ell+N(n)+1<0$, hence they are zero by
analytic continuation in $B$.] Using the expression of the
coefficients (\ref{alphakl}), we are next led to the alternative
expression
\begin{eqnarray}\label{intdeltalres}
&&\int d^3\mathbf{x}\,r^B \,\hat{x}_L\,\int^1_{-1} dz~ \delta_\ell(z)
\,\overline{\tau}^{\mu\nu} =
\nonumber\\&&\qquad\qquad\quad\frac{c^4}{16\pi
G}\sum_{k=0}^{+\infty}\,B(B+2\ell+4k+1)\,\alpha_{k,\ell}
\left(\frac{d}{cdt}\right)^{\!2k}\!\!\int
d^3\mathbf{x}\,r^{B+2k-2}\,\hat{x}_L\,\overline{h}^{\mu\nu}\,.~~
\end{eqnarray}

A remarkable feature of this result, which is the basis of our new
expressions, is the presence of an \textit{explicit factor} $B$ in
front of the integral. The factor means that the result depends only
on the occurrence of \textit{poles}, $\propto 1/B^p$, in the boundary
of the integral at infinity: $r\rightarrow +\infty$ with
$t=\mathrm{const}$. At this stage it is useful to write down the
expressions of the moments $I_L$ and $J_L$ we obtain by substituting
(\ref{intdeltalres}) back into (\ref{ILJL}). These are
{\allowdisplaybreaks
\begin{subequations}\label{ILJLinf}\begin{eqnarray} I_L &=&
\frac{c^2}{16\pi
G}\,\mathop{\mathrm{FP}}_{B=0}\,B\,r_0^{-B}\,\sum_{k=0}^{+\infty}
\left\{(B+2\ell+4k+1) \,\alpha_{k,\ell}
\left(\frac{d}{cdt}\right)^{\!2k} \int
d^3\mathbf{x}\,r^{B+2k-2}\hat{x}_L \left(\overline{h}^{00}
+\overline{h}^{ii}\right) \right.\nonumber\\
&&\qquad\qquad\left. -\frac{4(2\ell+1)(B+2\ell+4k+3)}{(\ell+1)
(2\ell+3)}\,\alpha_{k,\ell+1} \left(\frac{d}{cdt}\right)^{\!2k+1} \int
d^3\mathbf{x}\,r^{B+2k-2}\hat{x}_{iL} \,\overline
h^{i0}\right.\nonumber\\ &&\qquad\qquad\left.
+\frac{2(2\ell+1)(B+2\ell+4k+5)}{(\ell+1)(\ell+2)(2\ell+5)}
\,\alpha_{k,\ell+2} \left(\frac{d}{cdt}\right)^{2k+2} \int
d^3\mathbf{x}\,r^{B+2k-2}\hat{x}_{ijL} \,\overline
h^{ij}\right\}\,,\label{ILinf}\nonumber\\\\ J_L &=& \frac{c^3}{16\pi
G}\mathop{\mathrm{FP}}_{B=0}B\,r_0^{-B} \varepsilon_{ab\langle
i_\ell}\sum_{k=0}^{+\infty}\left\{(B+2\ell+4k+1) \,\alpha_{k,\ell}
\left(\frac{d}{c dt}\right)^{\!2k}\!\!\int d^3
\mathbf{x}\,r^{B+2k-2}\hat{x}_{L-1\rangle a}
\,\overline{h}^{b0}\right.\nonumber\\
&&\qquad\quad\left. -\frac{(2\ell+1)(B+2\ell+4k+3)}{
(\ell+2)(2\ell+3)}\,\alpha_{k,\ell+1} \left(\frac{d}{c
dt}\right)^{\!2k+1}\int d^3\mathbf{x}\, r^{B+2k-2}\hat{x}_{L-1\rangle
ac}\,\overline{h}^{bc}\right\}\,.\label{JLinf}\nonumber\\
\end{eqnarray}\end{subequations}}\noindent

Let us proceed further. Thanks to the factor $B$ we can replace the
integration domain of Eq.~(\ref{intdeltalres}) by some outer domain of
the type $r>\mathcal{R}$, where $\mathcal{R}$ denotes some large
arbitrary constant radius. The integral over the inner domain
$r<\mathcal{R}$ is always zero in the limit $B\rightarrow 0$ because
the integrand is constructed from $\overline{\tau}^{\mu\nu}$, and we
are considering extended regular PN sources, without
singularities. Now, in the outer (but still near-zone) domain we can
replace the PN metric coefficients $\overline{h}^{\mu\nu}$ by the
expansion in increasing powers of $1/r$ of the PN-expanded metric,
which is identical to the multipolar expansion of the PN-expanded
metric. This is precisely the quantity which was already introduced in
Eq.~(\ref{match}) and denoted there
$\mathcal{M}\bigl(\overline{h}^{\mu\nu}\bigr)$. Hence we have
\begin{eqnarray}\label{intdeltalresR}
&&\int d^3\mathbf{x}\,r^B \,\hat{x}_L\,\int^1_{-1} dz~ \delta_\ell(z)
\,\overline{\tau}^{\mu\nu} = \nonumber\\&&\qquad\quad\frac{c^4}{16\pi
G}\sum_{k=0}^{+\infty}\,B(B+2\ell+4k+1)\,\alpha_{k,\ell}
\left(\frac{d}{cdt}\right)^{\!2k}\!\!\int_{r>\mathcal{R}}\!\!
d^3\mathbf{x}\,r^{B+2k-2}\,\hat{x}_L\,\mathcal{M}\bigl(
\overline{h}^{\mu\nu}\bigr)\,.~~
\end{eqnarray}
We want now to make use of a more explicit form of the far near-zone
expansion $\mathcal{M}\bigl(\overline{h}^{\mu\nu}\bigr)$, whose
general structure is known. It consists of terms proportional to
arbitrary powers of $1/r$, and multiplied by powers of the
\textit{logarithm} of $r$. More precisely,
\begin{equation}\label{Mhexp}
\mathcal{M}\bigl(\overline{h}^{\mu\nu}\bigr)(\mathbf{x},t)=\sum_{a,\,b}
\frac{(\ln r)^b}{r^a}\,\varphi_{a,b}^{\mu\nu}(\mathbf{n},t)\,,
\end{equation}
where $a$ can take any positive or negative integer values, and $b$
can be any positive integer: $a\in\mathbb{Z}$, $b\in\mathbb{N}$. The
coefficients $\varphi_{a,b}^{\mu\nu}$ depend on the unit direction
$\mathbf{n}\equiv\mathbf{x}/r$ and on the coordinate time $t$ (in the
harmonic coordinate system). The structure (\ref{Mhexp}) for the
multipolar expansion of the near-zone (PN-expanded) metric is a
consequence, \textit{via} the matching equation (\ref{match}), of the
corresponding result concerning the structure of the
\textit{near-zone} expansion of the MPM metric (\ref{hMPM}), which has
been proved in Ref.~\cite{BD86} by using the properties of the MPM
algorithm for the iteration of the metric. [Again we stress the fact
that the result has been proved at some \textit{arbitrary but finite}
post-Minkowskian order $n$; see Eq.~(5.4) in Ref.~\cite{BD86}. In the
present paper we assume, following \cite{B98mult}, that we are always
entitled to sum up formally the post-Minkowskian series, and to view a
result like (\ref{Mhexp}) as true in the sense of formal power
series.]

As a side remark (which is not essential for the following), note that
in Ref.~\cite{BD86} the near-zone expansion of the MPM metric was
viewed as an expansion in ``ascending'' powers of $r/cT$, namely it
was written in the form
\begin{equation}\label{MhexpNZ}
\overline{\mathcal{M}\left(h^{\mu\nu}\right)}(\mathbf{x},t)=\sum_{p,\,q}
r^p\,(\ln r)^q\,f_{p,q}^{\mu\nu}(\mathbf{n},t)\,,
\end{equation}
where $p\in\mathbb{Z}$ and $q\in\mathbb{N}$. Evidently, the expansions
(\ref{MhexpNZ}) and (\ref{Mhexp}) are equivalent. The only difference
is that (\ref{Mhexp}) was ordered in ascending powers of $1/r$,
\textit{i.e.} in ``descending'' powers of $r$. Actually, both
expansions are formal Laurent-type expansions, valid in some
intermediate range of radii: $ a < r < c\,T$. Their coefficients are
related by a simple re-ordering of the exponents of $r$,
\begin{equation}\label{fphi}
f_{p,q}^{\mu\nu}(\mathbf{n},t)=\varphi_{-p,q}^{\mu\nu}(\mathbf{n},t)\,.
\end{equation}
In the following, we shall use the notation
$\varphi_{a,b}^{\mu\nu}(\mathbf{n},t)$, corresponding to
Eq.~(\ref{Mhexp}), for the coefficients of these equivalent
expansions.

Inserting (\ref{Mhexp}) into (\ref{intdeltalresR}), we are therefore
led to the computation of the integral
\begin{equation}\label{intexp}
\int_{r>\mathcal{R}}\!\!d^3\mathbf{x}\,r^{B+2k-2}\,\hat{x}_L
\,\mathcal{M}\bigl(\overline{h}^{\mu\nu}\bigr)=
\sum_{a,\,b}\int_\mathcal{R}^{+\infty}dr\,r^{B+2k+\ell-a}\,(\ln
r)^b\int d\Omega\,\hat{n}_L\,\varphi_{a,b}^{\mu\nu}(\mathbf{n},t)\,,
\end{equation}
where $d\Omega$ is the solid angle element associated with the unit
direction $\mathbf{n}$ (and $\hat{n}_L\equiv \hat{x}_L/r^\ell$). The
radial integral can be trivially integrated by analytic continuation
in $B$, with result
\begin{equation}\label{radial}
\int_\mathcal{R}^{+\infty}dr\,r^{B+2k+\ell-a}\,(\ln
r)^b=-\left(\frac{d}{dB}\right)^b\left[\frac{
\mathcal{R}^{B+2k+\ell-a+1}}{B+2k+\ell-a+1}\right]\,.
\end{equation}
Remember that we are ultimately interested only in the analytic
continuation of such integrals down to $B=0$. And as an integral such
as (\ref{radial}) is multiplied by a coefficient which is proportional
to $B$, we must control the poles of Eq.~(\ref{radial}) at
$B=0$. Those poles are in general multiple because of the presence of
powers of $\ln r$ in the expansion, and the consecutive multiple
differentiation with respect to $B$ in Eq.~(\ref{radial}). The poles
at $B=0$ clearly come from a single value of $a$, namely
$a=2k+\ell+1$. For that value, the ``multiplicity'' of the pole takes
the value $b + 1$. Here a useful simplification comes from the fact
that the factor in front of the integrals in (\ref{intdeltalresR}) is
of the form $\sim B(B+K)$. In other words, this factor contains only
the first and second powers of $B$. Therefore, only the simple and
double poles $1/B$ and $1/B^2$ in (\ref{radial}) can contribute to the
final result. Hence, we conclude that it is enough to consider the
values $b=0,1$ for the exponent $b$ of $\ln r$ in the expansion
(\ref{Mhexp}).

To express the result in the most convenient manner let us introduce a
special notation for some relevant combination of far-near-zone
coefficients $\varphi_{a,b}^{\mu\nu}(\mathbf{n},t)$, which as we just
said correspond exclusively to the values $a=\ell+2k+1$ and $b=0$ or
$1$. Namely,
\begin{equation}\label{Psi}
\Psi_{k,\ell}^{\mu\nu}(\mathbf{n},t)\equiv
\alpha_{k,\ell}\,\Bigl[-(2\ell+4k+1)\varphi_{2k+\ell+1,
0}^{\mu\nu}(\mathbf{n},t) +\Bigl(1-(2\ell+4k+1)\ln
r_0\Bigr)\varphi_{2k+\ell+1,1}^{\mu\nu}(\mathbf{n},t)\Bigr]\,,
\end{equation}
in which we have absorbed the numerical coefficient $\alpha_{k,\ell}$
defined by (\ref{alphakl}). [Notice that the coefficients
$\varphi_{a,b}^{\mu\nu}$ depend \textit{a priori} on the scale $r_0$.]
With this notation we then obtain
\begin{equation}\label{intexpres}
\mathop{\mathrm{FP}}_{B=0}\,B\,r_0^{-B}\,(B+2\ell+4k+1)\,\alpha_{k,\ell}
\int_{r>\mathcal{R}}\!\!d^3\mathbf{x}\,r^{B+2k-2}\,\hat{x}_L
\,\mathcal{M}\bigl(\overline{h}^{\mu\nu}\bigr)=4\pi
\left\langle\,\hat{n}_L\,\Psi_{k,\ell}^{\mu\nu}\right\rangle\,,
\end{equation}
where the brackets refer to the spherical or angular average (at
coordinate time $t$), \textit{i.e.}
\begin{equation}\label{angle}
\left\langle\,\hat{n}_L\,\Psi_{k,\ell}^{\mu\nu}\right\rangle (t)
\equiv\int
\frac{d\Omega}{4\pi}\,\hat{n}_L\,\Psi_{k,\ell}^{\mu\nu}(\mathbf{n},t)\,.
\end{equation}
As we can see, any reference to the intermediate scale $\mathcal{R}$
has completely disappeared. The quantities (\ref{angle}) are integrals
over a unit sphere, and can rightly be referred to as ``surface
integrals''. These surface integrals will be the basic blocks entering
our new expressions for the multipole moments. If we wish to
physically think of them as integrals over some two-surface
surrounding the source, we can roughly consider that this two-surface
is located at a radius $\mathcal{R}$, with $a \ll \mathcal{R} \ll
c\,T$. Anyway, the important point is that, as we have just remarked,
the surface integrals (\ref{angle}), and therefore the multipole
moments, are strictly independent of the choice of the intermediate
scale $\mathcal{R}$ which entered our reasoning.

Finally, we are in a position to write down the following final
results for the source multipole moments (\ref{ILinf}) and
(\ref{JLinf}), expressed solely in terms of the surface integrals of
the type (\ref{angle}), {\allowdisplaybreaks
\begin{subequations}\label{IJLfinal}\begin{eqnarray}
I_L &=& \frac{c^2}{4\,G}\,\sum_{k=0}^{+\infty}\biggl\{
\left(\frac{d}{cdt}\right)^{\!\!2k} \left\langle\,\hat{n}_L
\left(\Psi_{k,\ell}^{00}+\Psi_{k,\ell}^{ii}\right)\right\rangle\nonumber\\
&&\qquad\qquad -\frac{4(2\ell+1)}{(\ell+1)(2\ell+3)}
\,\left(\frac{d}{cdt}\right)^{\!\!2k+1} \left\langle\,\hat{n}_{iL}
\,\Psi_{k,\ell+1}^{i0}\right\rangle\nonumber\\ &&\qquad\qquad
+\frac{2(2\ell+1)}{(\ell+1)(\ell+2)(2\ell+5)}\,
\left(\frac{d}{cdt}\right)^{\!\!2k+2} \left\langle\,\hat{n}_{ijL}
\,\Psi_{k,\ell+2}^{ij}\right\rangle\biggr\}\,,\label{ILfinal}\\ J_L
&=& \frac{c^3}{4\,G}\,\varepsilon_{ab\langle
i_\ell}\sum_{k=0}^{+\infty}\biggl\{
\left(\frac{d}{cdt}\right)^{\!\!2k} \left\langle\,\hat{n}_{L-1\rangle
a}\,\Psi_{k,\ell}^{b0}\right\rangle\nonumber\\ &&\qquad\qquad
-\frac{2\ell+1}{(\ell+2)(2\ell+3)}
\,\left(\frac{d}{cdt}\right)^{\!\!2k+1}
\left\langle\,\hat{n}_{L-1\rangle ac}
\,\Psi_{k,\ell+1}^{bc}\right\rangle\biggr\}\label{JLfinal}\,.
\end{eqnarray}\end{subequations}}\noindent
The other source moments, $W_L$, $X_L$, $Y_L$ and $Z_L$, which
parametrize the gauge vector given by Eq.~(\ref{phi1}), admit similar
expressions, which can be derived by the same method. For these we
give only the results: {\allowdisplaybreaks
\begin{subequations}\label{WZLfinal}\begin{eqnarray} W_L &=&
\frac{c^3}{4\,G}\,\sum_{k=0}^{+\infty}\biggl\{
\frac{2\ell+1}{(\ell+1)(2\ell+3)}\,\left(\frac{d}{cdt}\right)^{\!\!2k}
\left\langle\,\hat{n}_{iL}\Psi_{k,\ell+1}^{i0}\right\rangle
\nonumber\\ &&\qquad\qquad -\frac{2\ell+1}{2(\ell+1)(\ell+2)(2\ell+5)}
\,\left(\frac{d}{cdt}\right)^{\!\!2k+1} \left\langle\,\hat{n}_{ijL}
\,\Psi_{k,\ell+2}^{ij}\right\rangle\biggr\}\,,\\ X_L &=&
\frac{c^4}{4\,G}\,\sum_{k=0}^{+\infty}\biggl\{
\frac{2\ell+1}{2(\ell+1)(\ell+2)(2\ell+5)}\,\left(\frac{d}{cdt}\right)^{\!\!2k}
\left\langle\,\hat{n}_{ijL}\Psi_{k,\ell+2}^{ij}\right\rangle\biggr\}\,,\\
Y_L &=&
\frac{c^4}{4\,G}\,\sum_{k=0}^{+\infty}\biggl\{-\left(\frac{d}{cdt}\right)^{\!\!2k}
\left\langle\,\hat{n}_L\Psi_{k,\ell}^{ii}\right\rangle\nonumber\\
&&\qquad\qquad +\frac{3(2\ell+1)}{(\ell+1)(2\ell+3)}
\,\left(\frac{d}{cdt}\right)^{\!\!2k+1}
\left\langle\,\hat{n}_{iL}\Psi_{k,\ell+1}^{i0}\right\rangle\nonumber\\
&&\qquad\qquad -\frac{2(2\ell+1)}{(\ell+1)(\ell+2)(2\ell+5)}
\,\left(\frac{d}{cdt}\right)^{\!\!2k+2} \left\langle\,\hat{n}_{ijL}
\,\Psi_{k,\ell+2}^{ij}\right\rangle\biggr\}\,,\\ Z_L &=&
\frac{c^4}{4\,G}\,\varepsilon_{ab\langle
i_\ell}\sum_{k=0}^{+\infty}\biggl\{ \frac{2\ell+1}{(\ell+2)(2\ell+3)}
\,\left(\frac{d}{cdt}\right)^{\!\!2k}
\left\langle\,\hat{n}_{L-1\rangle
ac}\,\Psi_{k,\ell+1}^{bc}\right\rangle\biggr\}\,.
\end{eqnarray}\end{subequations}}\noindent

\section{Application to a boosted Schwarzschild solution}\label{schw}

\subsection{Source quadrupole moment of the BSS at 3PN order}\label{sourcemoment}

As an application of our explicit surface-integral formulas
(\ref{IJLfinal}), we wish to compute the source-type multipole moments
of a spherically symmetric extended body moving with \textit{uniform}
velocity. Remember that our formalism assumes, in principle, that we
are dealing with regular, weakly self-gravitating bodies. We expect,
because of the nice ``effacing properties'' of Einstein's theory
\cite{D83houches}, that our final physical results, especially when
they are expressed as surface integrals like in (\ref{IJLfinal}), can
be applied to more general sources, such as neutron stars or black
holes. Indeed, we are going to confirm this expectation in the
simplest possible case, that of an isolated spherically symmetric body
which is known, by Birkhoff's theorem, to generate a universal
exterior gravitational field, given by the Schwarzschild solution. We
shall therefore apply our formulas to a \textit{boosted Schwarzschild
solution} (BSS).  Actually, in order to justify our use of the BSS in
standard harmonic coordinates, we must dispose of a small
technicality.

This technicality concerns the non-uniqueness of harmonic coordinates
for the Schwarzschild solution, even under the assumption of
stationarity (in the rest frame) and spherical symmetry. Indeed, under
these assumptions, and starting from the usual Schwarzschild-Droste
radial coordinate, say $r_S$, the (rest frame) radial coordinate of
the most general harmonic coordinate system, say $r = k(r_S)$, must
satisfy the differential equation (see, \textit{e.g.}, Weinberg
\cite{Weinberg}, page 181)
\begin{equation}\label{harm}
\frac{d}{d r_S}\left[\left(r_S^2 - \frac{2 \,G M}{c^2}\,r_S\right)
\frac{d k}{d r_S}\right] = 2\,k \,.
\end{equation}
The ``standard'' solution of Eq.~(\ref{harm}), which is considered in
all textbooks such as \cite{Weinberg}, reads simply
\begin{equation}\label{standsol}
r = k^\mathrm{standard}(r_S) = r_S - \frac{G M}{c^2}\,.
\end{equation}
In the black hole case, the solution (\ref{standsol}) is the only one
which is regular on the horizon, \textit{i.e.} when $r_S = 2 G M/c^2$
[as will be clear from Eqs.~(\ref{decaysol}) below]. However, in the
case of the external metric of an extended spherically symmetric body,
regularity on the horizon is not a relevant issue. What is relevant is
that the solution of the \textit{external} problem (\ref{harm}) be
smoothly matched to a \textit{regular} solution of the corresponding
\textit{internal} problem. As usual, this matching determines a unique
solution everywhere. In general, this unique, everywhere regular,
solution will correspond, in the exterior of the body, to a particular
case of the general, two-parameter solution of the second-order
differential equation (\ref{harm}). The latter is of the form
\begin{equation}\label{gensol}
r = k^\mathrm{general}(r_S) = c_1 \left(r_S - \frac{G M}{c^2} \right)+
c_2\,k_2(r_S) \,,
\end{equation}
where $k_2(r_S)$ denotes the (uniquely defined) ``radially decaying
solution'' of Eq.~(\ref{harm}), and where $c_1$ and $c_2$ are two
integration constants.  Indeed, when considering the flat-space limit
of Eq.~(\ref{harm}), it is easily seen that there are two independent
solutions which behave, when $r_S \rightarrow +\infty$, as $r_S$ and
$r_S^{-2}$ respectively. An explicit expression for the decaying
solution is\,\footnote{Here
$$F(\alpha,\beta,\gamma,z)=1+\frac{\alpha\beta}{\gamma}\,
\frac{z}{1!}+\frac{\alpha(\alpha+1)
\beta(\beta+1)}{\gamma(\gamma+1)}\,\frac{z^2}{2!}+\cdots\,,$$ denotes
Gauss' hypergeometric function.}
\begin{subequations}\label{decaysol}\begin{eqnarray}
k_2(r_S) &=& \frac{1}{r_S^2} \,F\left(2,2,4,\frac{2 \,G M}{c^2
r_S}\right)\,,\\ F(2,2,4,z) &=&
-\frac{6}{z^2}\left[2+\frac{1}{z}\left(2-z\right)\ln
\left(1-z\right)\right]\,.
\end{eqnarray}\end{subequations}
We can always normalize $c_1$ to the value $c_1 = 1$. Then, with the
above definitions, $c_2$ has the dimension of a length cubed. By
considering in more detail the matching of the general solution of the
harmonically relaxed Einstein equations at the 2PN level (see,
\textit{e.g.}, the book by Fock \cite{Fock}, page 322), one easily
finds that the second integration constant is of the order of $c_2
\sim (G M/c^2)^2 \,a$, where $a$ denotes the radius of the extended
body under consideration. It is also easily checked that the constant
$c_2$ parametrizes, at the linearized order, a \textit{gauge vector}
$\varphi^i_1$ of the form $\varphi^i_1 \propto c_2 \,\partial_i
(1/r)$, and can thus be referred to as a ``gauge
parameter''. Comparing to the general multipole decomposition
(\ref{phi1}), we see that this gauge parameter $c_2$ corresponds to
the monopole ($\ell =0$) in the gauge multipole sequence $X_L$.

Contrarily to the multipole moments of \textit{stationary} sources,
which are geometric invariants (and can be expressed as surface
integrals on a sphere at spatial infinity), the ``source multipole
moments'' defined in Ref.~\cite{B98mult} (and re-expressed above as
surface integrals over spheres in some intermediate region, $ a \ll r
\ll c\,T$) are probably not geometric invariants. They are useful
intermediate constructs, which allow one to compute physically
invariant information, but their definition is linked to the choice of
harmonic coordinates covering the source. This means that the various
``gauge multipoles'' $W_L, X_L, Y_L, Z_L$ will influence, at some
non-linear order of the MPM iteration, the values of the two sequences
of ``physical multipoles'': $I_L, J_L$. Therefore, one should expect
that, at some non-linear order, the physical multipoles $I_L, J_L$ of
a boosted general, harmonic-coordinate spherically symmetric metric
will start to depend on the value of the gauge parameter $c_2$.

Here, we are only interested in computing the quadrupole moment
$I_{ij}$ of a boosted general spherically symmetric metric. We shall
see below that the index structure of $I_{ij}$ will be provided by the
STF tensor product of the boost velocity $V^i$ with itself, denoted
$V^{\langle i}V^{j\rangle}$ (assuming that the origin of the
coordinates is at the initial position of the center of symmetry of
the BSS). Therefore, any contribution to $I_{ij}$ coming from the
gauge parameter $c_2$ must contain, at least, the factors $c_2$ and
$V^{\langle i}V^{j\rangle}$, and also the total mass $M$. Taking into
account the dimensionality of $c_2 \sim (G M/c^2)^2 \,a$, which is
that of a length cubed, it is easily seen that there is no way to
generate such a contribution to $I_{ij}$. Therefore, we conclude that
the source quadrupole moment of a boosted general, harmonic-coordinate
spherically symmetric metric is strictly equal to the source
quadrupole moment of a boosted \textit{standard harmonic-coordinate}
Schwarzschild solution, obtained by setting $c_2 =0$ (and $c_1 =1$) in
(\ref{gensol}), \textit{i.e.}  by choosing the standard harmonic
radial coordinate (\ref{standsol}).

In the following, we shall therefore consider only such a boosted
Schwarzschild solution (BSS) in standard form. We shall sometimes
refer to the source of this solution as a black hole (though, strictly
speaking, one should always have in mind some extended spherical
star). For simplicity, we shall translate the origin of the coordinate
system so that it is located at the initial position of the black hole
at coordinate time $t=0$. With this choice of origin of the
coordinates all the current-type moments $J_L$ of the BSS are zero. We
shall concentrate our attention on the mass-type quadrupole moment
$I_{ij}$, that we shall compute at the 3PN order.

Le us denote by $x^\mu=(c\,t,\mathbf{x})$ the global reference frame,
in which the black hole is moving, and by $X^\mu=(c\,T,\mathbf{X})$
the rest frame of the black hole --- both $x^\mu$ and $X^\mu$ are
assumed to be harmonic coordinates. Let $x^i(t)$ be the rectilinear
and uniform trajectory of the (center of symmetry of the) BSS in the
global coordinates $x^\mu$, and $\mathbf{V}=(V^i)$ be the constant
coordinate velocity of the BSS,
\begin{equation}\label{V}
V^i\equiv \frac{d x^i(t)}{d t}\,.
\end{equation}
The rest frame $X^\mu$ is transformed from the global one $x^\mu$ by
the Lorentz boost (for simplicity we consider a pure Lorentz boost
without rotation of the spatial coordinates)
\begin{equation}\label{xboost}
x^\mu = \Lambda^\mu_{~\nu}(\mathbf{V})\,X^\nu\,,
\end{equation}
whose components are explicitly given by
\begin{subequations}\label{boost}\begin{eqnarray}
\Lambda^0_{~0}(\mathbf{V})&=& \gamma\,,\\
\Lambda^i_{~0}(\mathbf{V})&=&\Lambda^0_{~i}(\mathbf{V})
=\gamma\frac{V^i}{c}\,,\\ \Lambda^i_{~j}(\mathbf{V})&=&
\delta^i_j+\frac{\gamma^2}{\gamma+1}\frac{V^iV_j}{c^2}\,,
\\\hbox{with}\quad\gamma &\equiv&
\left(1-\frac{V^2}{c^2}\right)^{-1/2}\,.
\end{eqnarray}\end{subequations}
As explained above, we can assume that the metric of the BSS in the
rest frame $X^\mu$ takes the standard harmonic-coordinate
Schwarzschild expression, which we write in terms of the Gothic metric
deviation $H^{\mu\nu}$, satisfying $\partial_\nu H^{\mu\nu}=0$. [We
use the same conventions as in Eq.~(\ref{harmonic}), with upper case
letters referring to quantities associated with the BSS rest frame.]
Hence,
\begin{subequations}\label{Hmunu}\begin{eqnarray}
H^{00} &=&
1-\frac{\left(1+\frac{G\,M}{c^2\,R}\right)^3}{1-\frac{G\,M}{c^2\,R}}\,,\\
H^{i0} &=& 0\,,\\ H^{ij} &=&
-\frac{G^2\,M^2}{c^4\,R^2}\,N^iN^j\,,\label{Hij}
\end{eqnarray}\end{subequations}
where $M$ is the total mass, $R\equiv\vert\mathbf{X}\vert$ and
$N^i\equiv X^i/R$. A well-known feature of the Schwarzschild metric in
harmonic coordinates is that the spatial Gothic metric $H^{ij}$ is
made of a single quadratic-order term $\propto G^2$ as shown in
Eq.~(\ref{Hij}). The Gothic metric deviation transforms like a Lorentz
tensor so the metric of the BSS in the global frame $x^\mu$ reads as
\begin{equation}\label{hmunu}
h^{\mu\nu}(x) =
\Lambda^\mu_{~~\rho}\Lambda^\nu_{~~\sigma}\,H^{\rho\sigma}(\Lambda^{-1}x)\,,
\end{equation}
in which the rest-frame coordinates have been expressed by means of
the global ones, \textit{i.e.} $X^\mu(x)=(\Lambda^{-1})^\mu_{~\nu}
x^\nu$, where the inverse Lorentz transformation is given by
$(\Lambda^{-1})^\mu_{~\nu}(\mathbf{V})\equiv\Lambda^{~\mu}_\nu(\mathbf{V})
=\Lambda^\mu_{~\nu} (-\mathbf{V})$. In our explicit calculations (done
with the software Mathematica) we employ the BSS metric in exactly the
form given by Eq.~(\ref{hmunu}). The only problem is to derive the
explicit relations giving the rest-frame radial coordinate $R$ and the
unit direction $N^i$ as functions of their global-frame counterparts
$r$ and $n^i$, of the global coordinate time $t$, and of the boost
velocity $V^i$. For these relations we find
\begin{subequations}\label{RNi}\begin{eqnarray}
R&=&r\,\biggl[1+c^2(\gamma^2-1)\left(\frac{t}{r}\right)^2-2\gamma^2
(Vn)\left(\frac{t}{r}\right)+\gamma^2\frac{(Vn)^2}{c^2}\biggr]^{1/2}\,,\\
N^i &=&\frac{n^i-\gamma
V^i\left(\frac{t}{r}\right)+\frac{\gamma^2}{\gamma+1}\frac{V^i}{c^2}(Vn)
}{\biggl[1+c^2(\gamma^2-1) \left(\frac{t}{r}\right)^2-2\gamma^2
(Vn)\left(\frac{t}{r}\right)+\gamma^2\frac{(Vn)^2}{c^2}\biggr]^{1/2}}\,,
\end{eqnarray}\end{subequations}
where $(Vn)\equiv \mathbf{V}\cdot\mathbf{n} = V^jn^j$ is the usual
Euclidean scalar product. The latter formulation of the BSS metric,
Eqs.~(\ref{Hmunu})-(\ref{RNi}), is well adapted to our calculations
because we have to perform, when computing the source multipole
moments, an integration over the \textit{coordinate} three-dimensional
spatial slice $\mathbf{x}\in\mathbb{R}^3$, with \textit{coordinate}
time $t=\mathrm{const}$, and this is easily done using the explicit
relations (\ref{RNi}).

However, let us notice that the BSS metric (in standard harmonic
coordinates) is best formulated in a manifestly Lorentz covariant way
as follows:
\begin{equation}\label{BSScov}
h^{\mu\nu} =
\left(1-\frac{\left(1+\frac{G\,M}{c^2\,r_\perp}\right)^3}{1
-\frac{G\,M}{c^2\,r_\perp}}\right)u^\mu u^\nu -
\frac{G^2\,M^2}{c^4\,r_\perp^2} \,n^\mu n^\nu\,,
\end{equation}
where $u^\mu$ is the time-like unit four-velocity of the center of
symmetry of the BSS, where $n^\mu$ is the space-like unit vector
pointing from the BSS to the field point along the direction
\textit{orthogonal} (in a Minkowskian sense) to the world line of the
BSS, and where $r_\perp$ denotes the orthogonal distance to the world
line (square root of the interval). The expression (\ref{BSScov}) is
completely equivalent to (and more elegant than) the more
``coordinate-rooted'' formulation (\ref{Hmunu})-(\ref{RNi}). We shall
employ it in a future investigation \cite{BDEI04dr}.

We compute the quadrupole moment $I_{ij}$ of the BSS, following the
prescriptions defined by Eq.~(\ref{ILfinal}). To this end we first
expand $h^{\mu\nu}$ when $c\rightarrow +\infty$, taking into account
all the $c$'s present both in the expression of the rest frame metric
(\ref{Hmunu}) as well as those coming from the Lorentz transformation
(\ref{hmunu})-(\ref{RNi}). In this process the boost velocity
$\mathbf{V}$ is to be considered as a constant, ``spectator'',
vector. Note in passing that, in the present problem, the
characteristic size $a$ of the source at time $t$ is given by the
displacement from the origin, $a \sim V \,t$, where $V\equiv
\vert\mathbf{V}\vert$, while the near-zone corresponds to $r \ll
c\,t$. Therefore, the far near-zone, where we read off the multipole
moments as some combination of expansion coefficients
$\varphi_{a,b}^{\mu\nu}(\mathbf{n},t)$, is the domain $V \,t \ll r \ll
c\,t$. We have evidently to assume that $V\ll c$ for this region to
exist.

We then first get the near-zone (or PN) expansion of the BSS metric,
$\overline{h}^{\mu\nu}$, by expanding in inverse powers of $c$ up to
3PN order. Next we compute the multipolar (or far) re-expansion of
each of the PN coefficients when $r\rightarrow +\infty$ with
$t=\mathrm{const}$. In this way we obtain what we have denoted by
$\mathcal{M}\bigl(\overline{h}^{\mu\nu}\bigr)$ in
Eq.~(\ref{Mhexp}). In the BBS case it is evident that the far-zone
expansion (\ref{Mhexp}) involves simply some powers of $1/r$, without
any logarithm of $r$ [indeed, see \textit{e.g.} Eq.~(\ref{RNi})].

With $\mathcal{M}\bigl(\overline{h}^{\mu\nu}\bigr)$ in hand we have
the coefficients of the various powers of $1/r$, and we obtain thereby
the various quantities $\Psi_{k,\ell}^{\mu\nu}$ defined by
Eq.~(\ref{Psi}). It is then a simple matter to compute all the
required angular averages present in our formula (\ref{ILfinal}) and
to obtain the following 3PN mass quadrupole moment of the BSS (the
angular brackets surrounding indices referring to the STF projection),
\begin{eqnarray}\label{Iij}
I_{ij}^\mathrm{\,BSS} &=& m\,t^2 \,V^{\langle
i}V^{j\rangle}\left[1+\frac{9}{14}\,\frac{V^2}{c^2}
+\frac{83}{168}\,\frac{V^4}{c^4}
+\frac{507}{1232}\,\frac{V^6}{c^6}\right] \nonumber\\ &+&
\frac{4}{7}\,\frac{G^2\,M^3}{c^6}\,V^{\langle i}V^{j\rangle} +
\mathcal{O}\left(\frac{1}{c^8}\right)\,.
\end{eqnarray} 
The first term represents the standard Newtonian expression,
multiplied here by a bunch of relativistic corrections. (Recall that
we have chosen the origin of the coordinate system at the initial
location of the BSS at $t=0$.)

The last term in Eq.~(\ref{Iij}), with coefficient $\mathcal{C}=4/7$,
is the most interesting for our purpose. It is purely of 3PN order,
and it contains the first occurence of the gravitational constant $G$,
which therefore arises in the quadrupole of the BSS only at 3PN
order. This term is interesting because it corresponds to one of the
regularization ``ambiguities'', due to an incompleteness of Hadamard's
self-field regularization, which recently appeared in the calculation
of the mass-type quadrupole moment of inspiralling point particle
binaries at the 3PN order \cite{BIJ02,BI04mult}. The associated
ambiguity parameter was called $\zeta$, and was introduced as a factor
of 3PN terms in the quadrupole moment having the form $\sim
m_1^3\,v_1^{\langle i}v_1^{j\rangle}$ or $m_2^3\,v_2^{\langle
i}v_2^{j\rangle}$, where $m_1$ and $m_2$ are the two point masses, and
$v_1^i$, $v_2^i$ are their coordinate velocities.\,\footnote{See
Section X of Ref.~\cite{BIJ02} for the discussion on why and how to
introduce the ambiguity parameters in the 3PN quadrupole of point
particle binaries.} The parameter $\zeta$ represents the analogue of
the ``kinetic ambiguity'' parameter $\omega_k$ in the 3PN Hamiltonian
of compact binaries \cite{JaraS98,JaraS99}. It is now clear that
$\zeta$ can be determined from what we shall now call the \textit{BSS
limit} of a binary system, which consists of setting one of the masses
of the binary to be \textit{exactly zero}, say $m_2=0$.

We have computed the BSS limit of the 3PN mass-type quadrupole moment
of compact binaries computed for general binary's orbits in
Refs.~\cite{BIJ02,BI04mult}. We have also inserted for the position of
the first body $y_1^i=v_1^i\,t$ in order to conform with our choice
for the origin of the coordinates. In this way we obtain
\begin{eqnarray}\label{Iijbin}
I_{ij}^\mathrm{\,BSS\,limit} &=& m_1\,t^2 \,v_1^{\langle
i}v_1^{j\rangle}\left[1+\frac{9}{14}
\,\frac{v_1^2}{c^2}+\frac{83}{168}\,\frac{v_1^4}{c^4}
+\frac{507}{1232}\,\frac{v_1^6}{c^6}\right]\nonumber\\ &+&
\biggl(\frac{232}{63}+\frac{44}{3}\zeta\biggr)
\frac{G^2\,m_1^3}{c^6}\,v_1^{\langle i}v_1^{j\rangle} +
\mathcal{O}\left(\frac{1}{c^8}\right)\,.
\end{eqnarray}
The comparison of Eqs.~(\ref{Iijbin}) and (\ref{Iij}) reveals a
complete match between the two results if and only if we have the
expected agreement between the masses, \textit{i.e.} $M=m_1$, and the
velocities, $v_1^i=V^i$ (since the velocity of the body remaining
after taking the BSS limit should exactly take the boost velocity),
and the ambiguity constant $\zeta$ takes the \textit{unique} value
\begin{equation}\label{zetares}
\zeta = -\frac{7}{33}\,.
\end{equation} 

Our conclusion, therefore, is that the ambiguity parameter $\zeta$ is
uniquely determined by the BSS limit. Because of the close relation
between the BSS limit with Lorentz boosts, it is clear that $\zeta$ is
linked to the Lorentz-Poincar\'e invariance of the multipole moment
formalism of Ref.~\cite{B98mult} as applied to compact binary systems
in \cite{BIJ02,BI04mult}. This link strongly suggests that the
specific value (\ref{zetares}) represents the only one for which the
expression of the 3PN quadrupole moment is compatible with the
Poincar\'e symmetry. In other words the present calculation indicates
that the Poincar\'e invariance should correctly be incorporated into
the laws of transformation of the source-type multipole moments for
general extended PN sources as given by Eqs.~(\ref{ILJL}) or
(\ref{IJLfinal}), though we have not verified this directly.

Note that the situation regarding $\zeta$ is the same as for the
kinetic ambiguity parameter $\omega_k$ in the 3PN equations of motion,
whose value has also been uniquely fixed by imposing the
Lorentz-Poincar\'e invariance of the formalism
\cite{BF00,BFeom,DJSpoinc}. (The other ambiguities $\xi$ and $\kappa$
in the binary's 3PN mass quadrupole moment \cite{BIJ02,BI04mult}
parametrize some Galilean invariant terms which cannot be derived from
a Lorentz transformation. This can also be seen from the fact that
$\xi$ and $\kappa$ are in factor of some ``interacting'' terms,
depending on both masses $m_1$ and $m_2$, and are thus out of the
scope of the BSS limit.) 

Let us finally emphasize that we have obtained Eq.~(\ref{zetares})
here without using any regularization scheme for curing the
divergencies associated with the self field of point
particles. However, it has been shown in recent related works
\cite{BDEI04,BDEI04dr} that this value is precisely the one derived in
the problem of point particles binaries at 3PN order by means of the
dimensional self-field regularization, instead of Hadamard's
regularization. (This shows that dimensional regularization is able to
correctly keep track of the global Poincar\'e invariance of the
general relativistic description of isolated systems.)

\subsection{Retarded far-zone field of the BSS at 3PN order}
\label{radiativemoment}

In this Section we present, as a further confirmation of our result,
an alternative derivation of the crucial coefficient $\mathcal{C}=4/7$
in Eq.~(\ref{Iij}), and hence of the kinetic-type ambiguity parameter
$\zeta=-7/33$. This new derivation will be based on the expansion of
the BSS metric (\ref{Hmunu})-(\ref{hmunu}) at Minkowskian
\textit{future null infinity}, $r\rightarrow +\infty$ with $u\equiv
t-r/c=\mathrm{const}$. It relies on the complete identification of the
metric (\ref{Hmunu})-(\ref{hmunu}), considered in the limit where
$r\rightarrow +\infty$ with $u=\mathrm{const}$, with the
multipolar-post-Minkowskian (MPM) external metric of an isolated
system, as given by the quantity $\mathcal{M}(h^{\mu\nu})$ reviewed in
Section \ref{recall} above. This identification is justified by the
fact that the general harmonic-coordinate MPM exterior metric defined
by Eq.~(\ref{hMPM}) smoothly matches, \textit{via} the matching
equation (\ref{match}), the inner PN solution of the inhomogeneous
field equations inside the matter source (see
Ref.~\cite{B98mult}). Indeed, so does the BSS metric
(\ref{Hmunu})-(\ref{hmunu}), as we have seen during the discussion on
the general solution of the harmonic-coordinates condition
(\ref{harm}) for the Schwarzschild metric, and our proof that the
parameter $c_2$ does not contribute to the quadrupole moment. 

In this Section we thus have to consider the BSS metric as a
functional of the \textit{six} sets of multipole moments
$\{I_L,\,J_L,\,W_L,\,X_L,\,Y_L,\,Z_L\}$ parametrizing the linearized
approximation to the MPM metric, namely $h_1^{\mu\nu}$ defined by
Eqs.~(\ref{h1MPM})-(\ref{phi1}), and then take into account the
subsequent non-linear iterations,
$h_2^{\mu\nu},\,h_3^{\mu\nu},\,\cdots$, whose computation is defined
by the MPM algorithm of Ref.~\cite{BD86}, as reviewed in
Eqs.~(\ref{hn})-(\ref{un}) above. Actually we shall work mostly with
the so-called ``\textit{canonical}'' MPM metric,
$h^{\mu\nu}_\mathrm{can}$, a functional of \textit{two and only two}
types of multipole moments called ``canonical'', mass-type moment
$M_L$ and current-type $S_L$, instead of the six source moments
$I_L,\,J_L,\,\cdots,\,Z_L$. Working with the metric
$h^{\mu\nu}_\mathrm{can}$ simplifies drastically the computation of
the non-linear interactions. We shall justify below why, for our
purpose, we can indeed use just this special case of a particular
harmonic-gauge-fixed external metric, rather than the more general
harmonic-coordinate external metric. Those metrics are geometrically
equivalent, but we shall have to check (as in the case of the $c_2$
parameter in Section \ref{sourcemoment}) that gauge effects do not
modify the result we are interested in.
 
In the present Section we shall have to compute the quadratic and
cubic metric corrections, $h^{\mu\nu}_{\mathrm{can}\,2}$ and
$h^{\mu\nu}_{\mathrm{can}\,3}$ in the notation of Eq.~(\ref{hMPM}),
for some specific multipole interactions. The canonical MPM metric is
defined by
\begin{equation}\label{hcan}
h^{\mu\nu}_\mathrm{can}[M_L,\,S_L] = \sum_{n=1}^{+\infty}
G^n\,h^{\mu\nu}_{\mathrm{can}\,n}\,,
\end{equation}
where the linearized approximation is given by the same formulas as
Eqs.~(\ref{h1can}) but with the canonical moments $M_L$ and $S_L$ in
place of the source moments $I_L$ and $J_L$, and with all the gauge
multipoles $\{W_L,\,X_L,\,Y_L,\,Z_L\}$ set to zero. In other words,
{\allowdisplaybreaks
\begin{subequations}\label{h1canMS}\begin{eqnarray}
h^{00}_{\mathrm{can}\,1}[M_L,\,S_L] &=& -\frac{4}{c^2}\sum_{\ell =
0}^{+\infty} \frac{(-)^\ell}{\ell !}  \partial_L \left[ \frac{1}{r}
M_L \right]\,,\label{h1canMS00}\\ h^{0i}_{\mathrm{can}\,1}[M_L,\,S_L]
&=& \frac{4}{c^3}\sum_{\ell = 1}^{+\infty} \frac{(-)^\ell}{\ell !}
\left\{ \partial_{L-1} \left[ \frac{1}{r} \dot{M}_{iL-1} \right] +
\frac{\ell}{\ell+1} \varepsilon_{iab} \,\partial_{aL-1} \left[
\frac{1}{r} S_{bL-1} \right]\right\}\,,\\
h^{ij}_{\mathrm{can}\,1}[M_L,\,S_L] &=&-\frac{4}{c^4}\sum_{\ell =
2}^{+\infty} \frac{(-)^\ell}{\ell !}  \left\{ \partial_{L-2} \left[
\frac{1}{r}\ddot{M}_{ijL-2} \right] + \frac{2\ell}{\ell+1}
\partial_{aL-2} \left[ \frac{1}{r} \varepsilon_{ab(i} \dot{S}_{j)bL-2}
\right]\right\}\,.\qquad
\end{eqnarray}
\end{subequations}}\noindent
Then the non-linear metrics $h^{\mu\nu}_{\mathrm{can}\,n}[M_L,\,S_L]$,
for $n\geq 2$, are obtained from (\ref{h1canMS}) by means of the same
algorithm as before, explained in Eqs.~(\ref{hn})-(\ref{un}), in the
case where the gauge vector $\varphi_1^\mu=0$, together with the
replacement $\{I_L,\,J_L\}\rightarrow\{M_L,\,S_L\}$.

We have already alluded to the important point, proved in
Ref.~\cite{BD86}, that although the canonical metric
(\ref{hcan})-(\ref{h1canMS}) is simpler than our previous construction
(\ref{hMPM})-(\ref{phi1}), it is physically or geometrically
equivalent to it, \textit{i.e.}  it describes the same physical matter
system, provided that $M_L$ and $S_L$ are related to
$I_L,\,J_L,\,\cdots,\,Z_L$ by some specific relations of the type
\begin{subequations}\label{MLSL}\begin{eqnarray}
M_L &=& I_L +
\mathcal{F}_L\left[I,\,J,\,W,\,X,\,Y,\,Z\right]\,,\label{MLIL}\\ S_L
&=& J_L + \mathcal{G}_L\left[I,\,J,\,W,\,X,\,Y,\,Z\right]\,,
\end{eqnarray}\end{subequations}
where $\mathcal{F}_L$ and $\mathcal{G}_L$ denote two non-linear
functionals (at least quadratic in the moments) of the original set of
source moments $I_L,\,J_L,\,\cdots,\,Z_L$. However, if the use of only
two sets of moments, $M_L$ and $S_L$, is very useful when computing
the non-linear multipole interactions, it remains that these moments
have still to be related to the more ``fundamental'' source moments by
Eqs.~(\ref{MLSL}). Indeed we know the analytic closed-form expressions
of the source moments $I_L,\,J_L,\,\cdots,\,Z_L$ (see Section
\ref{recall}) but similar formulas for $M_L,\,S_L$, valid to all PN
orders, are not known to exist. The equations (\ref{MLSL}) need to be
investigated anew for each specific cases. Fortunately, $M_L$ and
$S_L$ are ``almost'' equal to their counterparts $I_L$ and
$J_L$. Indeed we know that in the case of the mass-type moments for
instance, Eq.~(\ref{MLIL}) when further PN expanded reads as
\begin{equation}\label{MLPN}
M_L = I_L +\frac{1}{c^5}\delta I_L +
\mathcal{O}\left(\frac{1}{c^7}\right)\,,
\end{equation}
where $\delta I_L$ denotes some correction term arising at order 2.5PN
only (if necessary this term is given by Eq.~(4.24) in
Ref.~\cite{B96}). The remainder in Eq.~(\ref{MLPN}) is of order
3.5PN. The equation (\ref{MLPN}) shows that the 3PN term in the
quadrupole moment of the BSS we are looking for --- last term in
Eq.~(\ref{Iij}) --- will be \textit{the same} for $I_{ij}$ as for
$M_{ij}$. In addition to the relations (\ref{MLSL})-(\ref{MLPN}) we
must also take into account the possible effect of the coordinate
transformation between the canonical metric (\ref{hcan}) and the
metric (\ref{hMPM}), since as we mentioned, it is the latter which
must be identified with the retarded far-zone expansion of the BSS
metric (\ref{hmunu}).

Let us consider the general structure of the mass-type moments $M_L$
(or $I_L$) in the case of the BSS. In the present problem there is
only one vector which can be used to build the moment: Namely the
boost velocity $V^i$, so the index structure of $M_L$ must necessarily
be made of the STF product $V_{\langle L\rangle}\equiv V_{\langle
i_1}\cdots V_{i_\ell\rangle}$. [Indeed we recall that we choose the
origin of the coordinate system to lie on the trajectory of the BSS,
so we do not have at our disposal the vectorial separation between the
origin and the BSS world line. It is clear that such a restriction is
not physically crucial, however it simplifies the presentation and
several arguments in this Section very much.] In addition, we readily
see on dimensional grounds that $V_{\langle L\rangle}$ must be
multiplied by either $u^\ell$, where $u\equiv t-r/c$, or by the
product $(G\,M/c^3)\,u^{\ell-1}$, or by $(G\,M/c^3)^2\,u^{\ell-2}$,
and so on, and that each of the latter terms can be multiplied by some
relativistic corrections of the type $(V^2/c^2)^n$ up to any PN
order. Here $M$ denotes the BSS constant mass monopole, $\ell=0$, or
ADM mass. The general structure of the mass-type moment $M_L$ (and
also of $I_L$ as well) at the 3PN order therefore reads 
\begin{eqnarray}\label{MLstruct}
M_L(u) &=& M\,V_{\langle L\rangle}\left[u^\ell\left(1\,\,\&\,\,
\frac{V^2}{c^2}\,\,\&\,\,\frac{V^4}{c^4}\,\,\&\,\,\frac{V^6}{c^6}\right)
\,\,\&\,\,u^{\ell-1}
\frac{G\,M}{c^3}\left(1\,\,\&\,\,\frac{V^2}{c^2}\right)\right.\nonumber\\&&
\quad\qquad\left.\,\,\&\,\,u^{\ell-2}\left(\frac{G\,M}{c^3}\right)^2
+\mathcal{O}\left(\frac{1}{c^7}\right)\right]\,,
\end{eqnarray}
where the notation $\&$ means that we have to add a term having the
structure that is indicated next. In the quadrupole case, $\ell=2$, we
recognize in the last (explicit) term of Eq.~(\ref{MLstruct}) the
interesting form of the contribution to the ambiguity $\zeta$ in the
BSS limit.

The mass multipole moment of the BSS varies with time typically like
some $u^\ell$. This fact seems to be incompatible with the
construction of MPM metrics in Ref.~\cite{BD86}, since it was assumed
there, in order to implement this construction, that the matter source
is stationary before some fixed finite instant $-\mathcal{T}$ in the
remote past (and also that the coordinates are mass-centered in the
sense that the dipole moment $M_i$ is always zero). These assumptions,
made in Ref.~\cite{BD86} for purely technical reasons, imply in
principle that all the multipole moments $M_L$ are constant before the
date $-\mathcal{T}$ (\textit{i.e.} for $u < -\mathcal{T}$), and thus
cannot \textit{a priori} be applied to the physical situation of the
BSS. Nevertheless, we shall admit in the present paper that we are
allowed to use the construction and the results of \cite{BD86} even in
the case of the BSS. Indeed there has been several indications in our
previous works using the MPM formalism, notably when the formalism was
used for the computation of gravitational wave tails and
``tails-of-tails'' \cite{BD92,B98tail}, that the MPM expansion can in
fact be applied to more general sources which have always been
non-stationary, for instance an inspiralling compact binary formed by
capture of two particles initially moving on some hyperbolic-like
orbits. On the other hand, as we shall see our application of the MPM
formalism to the external field of the BSS will yield some consistent
result, which is independent of any initial instant $-\mathcal{T}$ and
is in agreement with the result of Section \ref{sourcemoment}. This
justifies \textit{a posteriori} (to some extent) our expectation that
the MPM formalism is still valid in the case of the BSS. Furthermore,
we shall give in Appendix \ref{nonlinear} an explicit proof that the
integration formulas of the MPM formalism admit a well-defined limit
when the multipole moments are continuously deformed into those of the
BSS.

We are interested in the quadrupolar contribution in the full
non-linearity expansion (\ref{hcan}), as seen from retarded infinity,
$r\rightarrow +\infty$ with $u=\mathrm{const}$ (this limit will be
referred below to as $\mathcal{I}^+$). We base our investigation on
the time-time component ($00$) of the metric, because this is that
component which is necessary and sufficient in order to obtain the
multipole moment itself, as opposed to some time derivative of it as
would be deduced from the $0i$ and $ij$ components (this is very
important because we are looking for a term in $M_{ij}$ which is a
constant). At the linearized approximation, the ``far-field''
quadrupole moment as seen from $\mathcal{I}^+$ simply reduces to the
canonical moment $M_{ij}$, and from Eq.~(\ref{h1canMS00}) we get (with
$\hat{n}^{ij}\equiv n^in^j-\frac{1}{3}\delta^{ij}$)
\begin{equation}\label{h00lindots}
G\,h^{00}_{\mathrm{can}\,1} = \cdots
-\frac{6\,G}{c^2\,r^3}\,\hat{n}^{ij}\,M_{ij}(u) + \cdots\,.
\end{equation}
Here we focus on the term of the form $\hat{n}^{ij}\,r^{-3}f(u)$, and
the ellipsis refer to all the other terms, either involving some other
multipolarities $\hat{n}_L$ with $\ell\not= 2$, or a power of $1/r$
different from 3. Consider now the corrections brought about by the
non-linear terms to be added to the linearized expression
(\ref{h00lindots}), and write
\begin{eqnarray}\label{h00candots}
h^{00}_{\mathrm{can}} &\equiv& \sum_{n=1}^{+\infty}
G^n\,h^{00}_{\mathrm{can}\,n} \nonumber\\&=& \cdots
-\frac{6\,G}{c^2\,r^3}\,\hat{n}^{ij}\,M_{ij}^\mathrm{\,rad}(u) +
\cdots\,,
\end{eqnarray}
where the far-field or ``radiative'' quadrupole moment (\textit{i.e.},
as seen from $\mathcal{I}^+$) is denoted by
$M_{ij}^\mathrm{\,rad}$. The non-linear terms in (\ref{h00candots})
introduce many couplings between the different multipole moments, and
there are a lot of possibilities in the general case. However, things
are much simpler in the case of the BSS owing to the particular
structure of the moments as determined in
Eq.~(\ref{MLstruct}). Notably, the $\ell$-th time derivative of the
BSS moment $M_L$ is always a constant. Then we find that at 3PN order
the \textit{most general} form of the allowed non-linear terms in
$M_{ij}^\mathrm{\,rad}$ reads as
\begin{eqnarray}\label{Mijrad}
M_{ij}^\mathrm{\,rad} = M_{ij}&+&\gamma \,\frac{G
\,M}{c^3}\dot{M}_{ij}+\epsilon \,\frac{G}{c^3}\,M_{\langle
i}\dot{M}_{j\rangle}\nonumber\\
&+&\sigma\,\frac{G}{c^5}\,\dot{M}_{k\langle i}\ddot{M}_{j\rangle
k}+\phi\,\frac{G}{c^5}\,M_k\dddot{M}_{ijk}
+\theta\,\frac{G}{c^5}\,\dot{M}_k\ddot{M}_{ijk}\nonumber\\ &+&\rho
\,\frac{G^2 \,M^2}{c^6}\ddot{M}_{ij}+\eta \,\frac{G^2
\,M}{c^6}\,\dot{M}_{\langle
i}\dot{M}_{j\rangle}+\mathcal{O}\left(\frac{1}{c^7}\right)\,,
\end{eqnarray}
where $\gamma,\,\epsilon,\,\sigma,\,\phi,\,\theta,\,\rho,\,\eta$
represent some unknown (for the moment at least) coefficients, which
are in general constant but as we shall see which can also depend on
the logarithm of the distance $r$. Note that Eq.~(\ref{Mijrad}) might
\textit{a priori} contain also some non-local contributions, say
$\int_{-\infty}^u dv \,M\ddot{M}_{ij}(v)$, or (worse)
$\int_{-\infty}^u dv \,M^2\dddot{M}_{ij}(v)$, but we shall discuss
such terms in Appendix \ref{nonlinear} and show that they do not
appear. The most important terms for the present purpose are the two
last ones, with coefficients $\rho$ and $\eta$, which involve the
\textit{cubic-order} multipole interactions $M^2\times M_{ij}$ and
$M\times M_i\times M_j$ (where $M$ is the mass and $M_i$ the mass
dipole of the BSS).

The coefficient we are looking for in the 3PN BSS source-type
quadrupole moment $I_{ij}$ was denoted by $\mathcal{C}$, and we now
write the corresponding term as
\begin{equation}\label{IijX}
\delta_\mathcal{C}I_{ij} = \mathcal{C}
\,\frac{G^2M^3}{c^6}\,V^{\langle i}V^{j\rangle}\,.
\end{equation}
The other terms in the BSS quadrupole have a different structure which
has already been displayed in Eq.~(\ref{Iij}). The result
$\mathcal{C}=4/7$ of Section \ref{sourcemoment} will be recovered by
the following method. As we noticed after Eq.~(\ref{MLPN}), the
coefficient $\mathcal{C}$ is necessarily the same for the source-type
and canonical-type moments, hence
\begin{equation}\label{MijX}
\delta_\mathcal{C}M_{ij} = \mathcal{C}
\,\frac{G^2M^3}{c^6}\,V^{\langle i}V^{j\rangle}\,.
\end{equation}
Consider next the radiative-type moment (\ref{Mijrad}), and look for
the modification of $\mathcal{C}$ induced by non-linearities. In order
to do this we have to remember the fact (see Ref.~\cite{B96}) that
both the canonical and source moments $M_L$ and $I_L$ for general
matter systems admit a PN expansion which is ``even'' up to 2PN order,
with the first ``odd'' correction being at the 2.5PN level. This
simple fact immediately shows that it is impossible that the ``odd''
terms shown in Eq.~(\ref{Mijrad}), which carry explicitly in front the
odd powers $1/c^3$ and $1/c^5$, contribute to a term at the 3PN
order. So we conclude that the sought-for modification of
$\mathcal{C}$ can come only from the two last terms, with coefficients
$\rho$ and $\eta$. Taking into account the Newtonian results
$\ddot{M}_{ij}=2 M\,V^{\langle i}V^{j\rangle}+\mathcal{O}(c^{-2})$ and
$\ddot{M}_i=M\,V^i+\mathcal{O}(c^{-2})$ we find
\begin{equation}\label{MijradX}
\delta_\mathcal{C}M_{ij}^\mathrm{\,rad} =
\left(\mathcal{C}+2\rho+\eta\right) \,\frac{G^2M^3}{c^6}\,V^{\langle
i}V^{j\rangle}\,.
\end{equation}
(Here our notation $\delta_\mathcal{C}M_{ij}^\mathrm{\,rad}$ means
that we are considering the complete term having the above indicated
structure.)

Next we come to the central part of this investigation, namely the
computation of the two coefficients $\rho$ and $\eta$. This task is
not so easy because $\rho$ and $\eta$ are in factor of some cubically
non-linear terms. We have obtained them by straightforward application
of the MPM algorithm of Refs.~\cite{BD86,B98quad}. Actually the value
of $\rho$, corresponding to the interaction $M^2\times M_{ij}$, is
already contained in the result of the calculation of the
gravitational wave ``tails-of-tails'' in Ref.~\cite{B98tail}. For
convenience we relegate the details of this non-linear iteration to
Appendix \ref{nonlinear}, and simply quote here our end results:
\begin{subequations}\label{deltaeta}\begin{eqnarray}
\rho&=&\frac{1271}{735}-\frac{58}{21}\ln\left(\frac{r}{r_0}\right)\,,\\
\eta&=&\frac{2918}{735}+\frac{116}{21}\ln\left(\frac{r}{r_0}\right)\,.
\end{eqnarray}\end{subequations}
For completeness we give also the known values of three other
constants in (\ref{Mijrad}):
\begin{equation}
\gamma=\frac{7}{2}\,,\qquad\epsilon=\frac{7}{3}\,,
\qquad\sigma=\frac{20}{21}\,.
\end{equation}
These values come from Eq.~(2.8a) in Ref.~\cite{B98tail} for $\gamma$,
Eq.~(\ref{A7a}) in Appendix A below for $\epsilon$, and the Table 2 in
Ref.~\cite{B98quad} for $\sigma$. (We have not computed the
coefficients $\phi$ and $\theta$.)

As we see from (\ref{deltaeta}), both $\rho$ and $\eta$ depend on
the logarithm of $r/r_0$, where $r_0$ is the same constant as in the
source multipole moments (\ref{ILJL}), but we nicely find that the
logarithms cancel out in the relevant combination of these
coefficients which enters Eq.~(\ref{MijradX}). In fact, we can argue
that the cancellation of the logarithms must necessarily occur because
nowhere in the far-zone expansion of the BSS metric
(\ref{Hmunu})-(\ref{hmunu}) can such logarithms of $r/r_0$ be
generated. Hence we get
\begin{equation}\label{MijradX'}
\delta_\mathcal{C}M_{ij}^\mathrm{\,rad} =
\left(\mathcal{C}+\frac{52}{7}\right) \,\frac{G^2M^3}{c^6}\,V^{\langle
i}V^{j\rangle}\,.
\end{equation}

Let us now check that there are no gauge effects, linked to the
non-geometrical nature of our definitions for the multipole moments,
concerning the particular term we consider in (\ref{MijradX'}), in the
sense that the coordinate transformation between the canonical metric
coefficient $h^{00}_\mathrm{can}$ and the corresponding BSS one,
computed from the expansion of Eqs.~(\ref{Hmunu})-(\ref{hmunu}) and
given by $\mathcal{M}(h^{00})$ in the MPM formalism, has no effect on
this particular term. The proof goes by noticing first that the
coordinate transformation at the linearized level is given by the
gauge vector (\ref{phi1}) parametrized by the four source-type moments
$W_L$, $X_L$, $Y_L$ and $Z_L$. The latter moments were given in
Eqs.~(5.17)-(5.20) of Ref.~\cite{B98mult}, and we have provided in
Eqs.~(\ref{WZLfinal}) above their new forms in terms of surface
integrals. An important point is that the moments
$W_L,\,X_L,\,Y_L,\,Z_L$ have been defined in such a way that they
admit some non-zero \textit{finite} limits when $c\rightarrow
+\infty$; in other words, they ``start at Newtonian order'' and their
Newtonian limit is non-zero. By using this fact together with
dimensional analysis, it is a simple matter to write down their
structures in the case of the BSS, in a manner similar to what we did
for the moment $M_L$ in Eq.~(\ref{MLstruct}). We find
\begin{subequations}\label{WXYstruc}\begin{eqnarray}
W_L &=& V^2\,u\,\Bigl\{\hbox{same structure as the one of $M_L$ given
by (\ref{MLstruct})}\Bigr\}\,,\\ X_L &=& V^4\,u^2\,\Bigl\{\hbox{same
structure}\Bigr\}\,,\\ Y_L &=& V^2\,\Bigl\{\hbox{same
structure}\Bigr\}\,,
\end{eqnarray}\end{subequations}
while $Z_L$ is a current-type moment so $Z_L=0$ with our choice of
origin for the BSS. The structures (\ref{WXYstruc}) imply that the
potentially dangerous term, which is proportional to $M^3$, must
necessarily appear in these moments at order $1/c^6$ relatively to the
Newtonian order. Next one readily shows, again on dimensional grounds,
that the only possible modifications of the quadrupole moment $M_{ij}$
in (\ref{h00lindots}) which are due to the gauge transformation, take
the forms $\dot{W}_{ij}/c^2$, $\ddot{X}_{ij}/c^4$ or $Y_{ij}/c^2$. It
is therefore impossible, because of the latter extra factors $1/c^2$
or $1/c^4$, that a dangerous term be generated in this way at the 3PN
order. Similar arguments are even more easily applied at non-linear
order in the coordinate transformation: For instance we find that it
is impossible that a non-linear coupling of the type $M_L\times W_P$,
or $W_L\times X_P$, has the correct structure at 3PN order in the
quadrupole moment.

This check being done, we conclude that the coefficient we predicted
for the relevant term in Eq.~(\ref{MijradX'}) represents exactly the
quadrupolar contribution in the retarded far-zone expansion (at
$\mathcal{I}^+$) of the BSS metric (and by the reasoning of Section
\ref{sourcemoment} we know that we can use the BSS metric in standard
harmonic coordinates). Now the point is that we can also compute
\textit{directly} this contribution in the far-zone expansion of the
BSS metric by using the formulas (\ref{Hmunu})-(\ref{hmunu}). Since
the far-zone expansion is to be done at retarded time (and not at time
$t=\mathrm{const}$ as we did in Section \ref{sourcemoment}), we must
for this calculation \textit{substitute $t$ by $u+r/c$} in
Eqs.~(\ref{RNi}) and only afterwards take the limit $r\rightarrow
+\infty$ (holding $u=\mathrm{const}$). In this way we obtain the
quadrupolar piece $\propto \hat{n}^{ij}/r^3$ in the 00 component of
the metric, and as we have proved before we are allowed to
\textit{identify} the term therein having the correct structure with
the one computed in (\ref{MijradX'}). A simple Mathematica calculation
reveals that the term in question has the coefficient: $8$. Therefore
$\mathcal{C}+\frac{52}{7}=8$ and we obtain
\begin{equation}\label{C}
\mathcal{C}=\frac{4}{7}\,,
\end{equation}
in complete agreement with our previous finding (\ref{Iij}), and in
support of the value for the kinetic ambiguity parameter:
$\zeta=-7/33$.

\appendix \section{Non-linear multipole interactions}\label{nonlinear}

This Appendix is devoted to the computation of the cubically
non-linear coefficients $\rho$ and $\eta$ entering
Eq.~(\ref{Mijrad}). Following the MPM algorithm of Ref.~\cite{BD86} we
must first compute the non-linear cubic source term, say
$\Lambda^{\mu\nu}_{\mathrm{can}\,3}$, which is composed of the sum of
a quadratic-order piece made out of products between
$h_{\mathrm{can}\,1}$ and $h_{\mathrm{can}\,2}$ (and of course their
gradients), and of a purely cubic-order piece, involving three factors
$h_{\mathrm{can}\,1}$. The linearized metric has been given in
(\ref{h1canMS}), and from it all subsequent iterations are generated
by the MPM algorithm, which is the same as in
Eqs.~(\ref{hn})-(\ref{un}) except that, for the ``canonical''
construction, the gauge vector is $\varphi_1^\mu=0$ and we use $M_L$
and $S_L$ as moments instead of $I_L,\,J_L$. The cubic source term
$\Lambda^{\mu\nu}_{\mathrm{can}\,3}
\left[h_{\mathrm{can}\,1},h_{\mathrm{can}\,2}\right]$ is inverted by
means of the retarded d'Alembertian operator, regularized by the
specific finite part $\mathop{\mathrm{FP}}_{B=0}$ of Eq.~(\ref{un}),
\begin{equation}\label{A1}
u^{\mu\nu}_{\mathrm{can}\,3}=\mathop{\mathrm{FP}}_{B=0}\Box^{-1}_\mathrm{R}
\biggl[\left(\frac{r}{r_0} \right)^B
\Lambda^{\mu\nu}_{\mathrm{can}\,3}\biggr]\,.
\end{equation}
The metric at cubic order reads then
\begin{equation}\label{A2}
h^{\mu\nu}_{\mathrm{can}\,3}=u^{\mu\nu}_{\mathrm{can}\,3}
+v^{\mu\nu}_{\mathrm{can}\,3}\,,
\end{equation}
where $v^{\mu\nu}_{\mathrm{can}\,3}$ represents a particular
homogeneous solution of the wave equation, such that the harmonicity
condition $\partial_\nu h^{\mu\nu}_{\mathrm{can}\,3}=0$ is
satisfied. Below we shall not need to compute
$v^{\mu\nu}_{\mathrm{can}\,3}$, since we shall simply have to invoke
the fact that its $00$ component, $v^{00}_{\mathrm{can}\,3}$, is made
of multipolarities $\ell=0$ or $1$ only (see \textit{e.g.} Eq.~(2.12a)
in \cite{B98quad}), and will thus always be zero for the quadrupole
case $\ell=2$ of concern to us here.

Consider first the cubic interaction $M\times M\times M_{ij}$. In this
case $\Lambda^{\mu\nu}_{\mathrm{can}\,3}$ is already known from
previous work, Eq.~(4.16a) of Ref.~\cite{B98tail}, which gives, for
the needed $00$ component,\,\footnote{In this Appendix we pose
$G=c=1$.}
\begin{equation}\label{A3}
\Lambda^{00}_{\mathrm{can}\,3}=\hat{n}_{ab}M^2\left[-516\,r^{-7}\,M_{ab}(u)
-516\,r^{-6}\,\dot{M}_{ab}(u)-304\,r^{-5}\,\ddot{M}_{ab}(u)\right]\,.
\end{equation}
To integrate we use the formulas given in the Appendix A of
Ref.~\cite{B98quad}. We limit ourselves to the computation of the part
$u^{\mu\nu}_{\mathrm{can}\,3}$ of the algorithm since
$v^{00}_{\mathrm{can}\,3}=0$ for the particular multipole interaction
we consider. In all this calculation we use the fact that the second
time derivative of the BSS quadrupole moment is constant, hence
$\dddot{M}_{ij}=0$. First of all, by straightforward use of the
integration formula (A.16a) of \cite{B98quad} we obtain
\begin{eqnarray}\label{A3'}
h^{00}_{\mathrm{can}\,3}&=&\hat{n}_{ab}M^2\left[-\frac{516}{14}\,r^{-5}
\,M_{ab}(u)-\frac{516}{14}\,r^{-4}\,\dot{M}_{ab}(u)-
\frac{3354}{245}\,r^{-3}\,\ddot{M}_{ab}(u)\right]\nonumber\\
&-&\frac{580}{7} \,\mathop{\mathrm{FP}}_{B=0}\Box^{-1}_\mathrm{R}
\biggl[\left(\frac{r}{r_0}\right)^B r^{-5}\,\hat{n}_{ab}
\,M^2\,\ddot{M}_{ab}(u)\biggr]\,.
\end{eqnarray}

The last term is \textit{a priori} more delicate in the case of the
BSS because we know from Eq.~(A.13) of \cite{B98quad} that it could
generate an integral depending on the whole time evolution of the
system, such as a ``tail'' integral of the type
$J(u)\equiv\int_{-\infty}^u dv \,M^2\dddot{M}_{ij}(v)$. The value of
an integral like $J$ would be quite ambiguous within our MPM-based
approach. Indeed, on the one hand, the boosted system that we finally
consider has $M_{ij}(u)$ proportional to $u^2$, so that
$\dddot{M}_{ij}(u)$ vanishes identically. We would therefore expect,
from this point of view, that $J$, being the integral of a vanishing
integrand, is zero: $J=0$. On the other hand, we can perform the
$v$-integral in $J$ to get $J=M^2\,
[\ddot{M}_{ij}(u)-\ddot{M}_{ij}(-\infty)]$. Now, as we recalled above,
the MPM framework on which we base our discussion assumes that we
start initially by considering systems such that the multipole moments
become time-independent in the remote past (before some finite instant
$-\mathcal{T}$). For such systems, the second contribution in the
latter expression for $J$ vanishes, and we would get:
$J=M^2\ddot{M}_{ij}(u)$, which does not vanish in the case of a
boosted source.

However, this ambiguous situation does not appear in the present
calculation. Indeed, the logic of our calculation is the following. To
derive the MPM result (\ref{A3'}) we had to initially assume that
$M_{ij}(u)$ tends fast enough toward a constant in the infinite
past. Then, after having done the MPM iteration we get the form
(\ref{A3'}). Now, starting from the explicit expression (\ref{A3'}) we
want to relax the original assumption about $M_{ij}(u)$, and consider
a \textit{deformation process} in which $M_{ij}(u)$ interpolates
between an initial MPM-like $M_{ij}(u)$ (tending to a constant in the
past) and a final $M_{ij}^\mathrm{BSS}(u)$ of the form of $u^2$. The
question is then to know (i) whether the R.H.S. of Eq.~(\ref{A3'})
admits any limit after this continuous deformation process, and (ii)
what is the value of this limit. In mathematical terms the question is
essentially a question of interchange of a limit with an integral
operation, \textit{i.e.} whether $\lim_n \int f_n(u)\,du = \int \lim_n
f_n(u)\,du$ holds, for a sequence of functions $f_n(u)$. As we know
the answer is positive under ``good'' conditions, for instance of
uniform convergence, or more generally (Lebesgue's theorem) of
dominated convergence, which says essentially that if $\vert
f_n(u)\vert < g(u)$ and if $\int g(u)\,du$ is finite, then we can
interchange the limits. In our case, we can use Lebesgue's theorem of
dominated convergence (see, \textit{e.g.}, the book
\cite{ChoquetDewitt}), which is both simple and powerful.

The only delicate term in the R.H.S. of (\ref{A3'}) is the last,
integral term. We must interpolate between some initial
$\ddot{M}_{ij}(u)$ which vanishes in the past to become a constant in
the future, and a final $\ddot{M}_{ij}^\mathrm{BSS}(u)$ which is
always constant. It is clear that we can do this interpolation in a
way that $\vert\ddot{M}_{ij}(u)\vert$ remains \textit{always
bounded}. Using such a bound in the last term of (\ref{A3'}), which is
a three-dimensional (retarded) integral, we easily see that, under the
assumption of a bounded $\vert\ddot{M}_{ij}(u)\vert$, the integrand is
bounded by a (positive) function which \textit{is integrable} (because
of the fast convergence brought by the $r^{-5}$ factor, together with
the $1/r$ factor contained in the propagator). Therefore, we can
indeed interchange the limiting process and the integration one, and
conclude that the limit of the L.H.S. of Eq.~(\ref{A3'}) exists, and
is simply given by replacing $M_{ij}(u)$ in the R.H.S. by its limiting
expression for a BSS which is: $M_{ij}^\mathrm{BSS}(u)$ proportional
to $u^2$. As, under this limit, the integrand of the last term in
(\ref{A3'}) becomes time-independent, we can explicitly compute the
limit by replacing the retarded propagator by a Poisson
integral. Hence, we get
\begin{eqnarray}\label{A3''}
\mathop{\mathrm{FP}}_{B=0}\Box^{-1}_\mathrm{R}
\biggl[\left(\frac{r}{r_0}\right)^B r^{-5}
\,\hat{n}_{ab}\,M^2\,\ddot{M}_{ab}\biggr] &=&
\mathop{\mathrm{FP}}_{B=0}\Delta^{-1}
\biggl[\left(\frac{r}{r_0}\right)^B
r^{-5}\,\hat{n}_{ab}\biggr]\,M^2\,\ddot{M}_{ab}\nonumber\\ &=&
-\frac{1}{5}\,r^{-5}\left[\frac{1}{5}
+\ln\left(\frac{r}{r_0}\right)\right]
\hat{n}_{ab}\,M^2\,\ddot{M}_{ab}\,.
\end{eqnarray}
One easily checks that the result (\ref{A3''}) agrees with the more
formal way of doing the calculation which consists of applying the
formula (A.13) of Ref.~\cite{B98quad} to the BSS case. As we can see,
the last term in the R.H.S. of the latter formula, which is given by a
``tail'' integral, vanishes when we insert the BSS quadrupole into the
source term of the retarded integral, since this source term involves
in fact the fifth time derivative of the BSS quadrupole which is
zero. On the other hand, the logarithmic term in the formula (A.13) of
\cite{B98quad} does remain, and then we recover exactly
Eq.~(\ref{A3''}) --- with the same constant $r_0$ on both sides of the
equation. The proof that we have detailed above, based on Lebesgue's
theorem of dominated convergence, rigorously justifies (for the case
at hand) that one can ``blindly'' use the formulas in the Appendix of
\cite{B98quad} even for the case of the BSS.

Finally, gathering Eqs.~(\ref{A3'}) and (\ref{A3''}) we obtain
\begin{eqnarray}\label{A4}
h^{00}_{\mathrm{can}\,3}&=&\hat{n}_{ab}M^2\left(-\frac{516}{14}r^{-5}
M_{ab}(u)-\frac{516}{14}r^{-4}\dot{M}_{ab}(u) +
\left[-\frac{2542}{245}+\frac{116}{7}
\ln\left(\frac{r}{r_0}\right)\right]r^{-3}\ddot{M}_{ab}\right)\,,\nonumber\\
\end{eqnarray}
which shows by comparing to (\ref{h00candots})-(\ref{Mijrad}) that the
coefficient we are seeking is
\begin{equation}\label{A5}
\rho=\frac{1271}{735}-\frac{58}{21}\ln\left(\frac{r}{r_0}\right)\,.
\end{equation}

The cubic multipole interaction $M\times M_i\times M_j$ is much longer
to obtain because we are obliged to compute it from scratch (no
earlier results in the literature are available). At linearized order
the metric reads
\begin{subequations}\label{A6}\begin{eqnarray}
h^{00}_{\mathrm{can}\,1}&=&-4\,r^{-1} M
+4\,\partial_a\left[r^{-1}\,M_a(u)\right]\,,\\
h^{i0}_{\mathrm{can}\,1}&=&- 4\,r^{-1}\dot{M}_i(u)\,,\\
h^{ij}_{\mathrm{can}\,1}&=&0\,.
\end{eqnarray}\end{subequations}
Straightforward calculations following the MPM algorithm then yield
the part of the metric at quadratic order corresponding to the
multipole couplings $M\times M_i$ and $M_i\times M_j$,
{\allowdisplaybreaks
\begin{subequations}\label{A7}\begin{eqnarray}
h^{00}_{\mathrm{can}\,2}&=&n_a
\,M\left(-14\,r^{-3}\,M_a-14\,r^{-2}\,\dot{M}_a\right)\nonumber\\
&+&\hat{n}_{ab}\left(-7\,r^{-4}\,M_a\,M_b-14\,r^{-3}
\,M_a\,\dot{M}_b-36\,r^{-2} \,\dot{M}_a\,\dot{M}_b\right)\nonumber\\
&+&\left(-\frac{7}{3}\,r^{-4}\,M_a\,M_a-\frac{14}{3}\,r^{-3}
\,M_a\,\dot{M}_a+\frac{23}{9}\,r^{-2}
\,\dot{M}_a\,\dot{M}_a\right)\,,\label{A7a}\\
h^{i0}_{\mathrm{can}\,2}&=&-\hat{n}_{ia}\,r^{-2}\,M
\,\dot{M}_a-\frac{22}{3}\,r^{-2}\,M\,\dot{M}_i\nonumber\\
&+&\hat{n}_{iab}\left(-2\,r^{-3}\,M_a\,\dot{M}_b-2\,r^{-2}
\,\dot{M}_a\,\dot{M}_b\right)\nonumber\\
&+&n_i\left(-\frac{17}{5}\,r^{-3}\,M_a\,\dot{M}_a-\frac{17}{5}\,r^{-2}
\,\dot{M}_a\,\dot{M}_a\right)\nonumber\\
&+&n_a\left(r^{-3}\left[-\frac{37}{5}\,\dot{M}_i\,M_a+
\frac{18}{5}\,M_i\,\dot{M}_a\right]-\frac{19}{5}\,r^{-2}
\,\dot{M}_i\,\dot{M}_a\right)\,,\\
h^{ij}_{\mathrm{can}\,2}&=&\hat{n}_{ija}
\,M\left(-4\,r^{-3}\,M_a-4\,r^{-2}\,
\dot{M}_a\right)\nonumber\\&+&\delta_{ij}\,n_a
\,M\left(-\frac{4}{5}\,r^{-3}\,M_a-\frac{4}{5}\,r^{-2}\,
\dot{M}_a\right)\nonumber\\&+&n_{(i}
\,M\left(\frac{2}{5}\,r^{-3}\,M_{j)}+\frac{2}{5}\,r^{-2}
\,\dot{M}_{j)}\right)\nonumber\\&+&\hat{n}_{ijab}\left(
-\frac{9}{2}\,r^{-4}\,M_a\,M_b-9\,r^{-3}\,M_a\,\dot{M}_b-4\,
r^{-2}\,\dot{M}_a\,\dot{M}_b\right)
\nonumber\\&+&\delta_{ij}\,\hat{n}_{ab}\left(
-\frac{1}{7}\,r^{-4}\,M_a\,M_b-\frac{2}{7}\,r^{-3}\,M_a \,\dot{M}_b
+\frac{24}{7}\,r^{-2}\,\dot{M}_a\,\dot{M}_b\right)\nonumber\\
&+&\hat{n}_{a(i}\left( -\frac{4}{7}\,r^{-4}\,M_a\,M_{j)}-
\frac{4}{7}\,r^{-3}\left[\dot{M}_a\,M_{j)}+M_a\,\dot{M}_{j)}\right]
-\frac{58}{7}\,r^{-2}\,\dot{M}_a\,\dot{M}_{j)}\right)\nonumber\\
&+&\hat{n}_{ij}\left(
\frac{6}{7}\,r^{-4}\,M_a\,M_a+\frac{12}{7}\,r^{-3}\,M_a\,\dot{M}_a
+\frac{31}{7}\,r^{-2}\,\dot{M}_a\,\dot{M}_a\right)\nonumber\\
&+&\delta_{ij}\left(
-\frac{2}{15}\,r^{-4}\,M_a\,M_a-\frac{4}{15}\,r^{-3}\,M_a\,\dot{M}_a
-\frac{19}{15}\,r^{-2}\,\dot{M}_a\,\dot{M}_a\right)\nonumber\\&+&
\frac{1}{15}\,r^{-4}\,M_i\,M_j+\frac{2}{15}\,r^{-3}\,M_{(i}\,\dot{M}_{j)}
-\frac{38}{15}\,r^{-2}\,\dot{M}_i\,\dot{M}_j\,.
\end{eqnarray}\end{subequations}}\noindent
Using such expressions (\ref{A6}) and (\ref{A7}) we next obtain the
source term at the cubic-order approximation $M\times M_i\times
M_j$. We are interested only in its $00$ component which is then found
to be
\begin{eqnarray}\label{A8}
\Lambda^{00}_{\mathrm{can}\,3} &=& \hat{n}_{ab} \left( -324\,r^{-7}\,M
\,M_a\,M_b-648\,r^{-6}\,M \,M_a\,\dot{M}_b
-112\,r^{-5}\,M\,\dot{M}_a\,\dot{M}_b\right)\nonumber\\&& -160\,r^{-7}\,M
\,M_a\,M_a-320\,r^{-6}\,M \,M_a\,\dot{M}_a -\frac{644}{9}\,r^{-5}\,M
\,\dot{M}_a\,\dot{M}_a\,.
\end{eqnarray}
The integration proceeds exactly in the same way as in
Eqs.~(\ref{A3'})-(\ref{A3''}). Again, the problem is that of
interchanging a limiting process
$\dot{M}_i(u)\longrightarrow\dot{M}_i^\mathrm{BSS}(u)$ with the
retarded integration, and this can be proved by using the Lebesgue
theorem, because the various powers of $1/r$ in the R.H.S. of
(\ref{A8}) ensure that, in a process where $\vert\dot{M}_i(u)\vert$
remains bounded, the integral is bounded by a positive convergent
integral. And again the result agrees with the one we would formally
obtain by using the formulas in the Appendix of \cite{B98quad}. As
before we have $v^{00}_{\mathrm{can}\,3}=0$ from the argument
concerning the multipolarity $\ell=0,1$ of this special piece. Our
result is then
\begin{eqnarray}\label{A9}
h^{00}_{\mathrm{can}\,3} &=& \hat{n}_{ab} \left(
-\frac{162}{7}\,r^{-5}\,M \,M_a\,M_b-\frac{324}{7}\,r^{-4}\,M
\,M_a\,\dot{M}_b\right.\nonumber\\&&\quad\quad\left.
+\left[-\frac{5836}{245}-\frac{232}{7}\ln\left(\frac{r}{r_0}\right)
  \right]\,r^{-3}\,M\,\dot{M}_a\,\dot{M}_b\right)\nonumber\\&& -
8\,r^{-5}\,M \,M_a\,M_a-16\,r^{-4}\,M \,M_a\,\dot{M}_a
+\frac{110}{27}\,r^{-3}\,M \,\dot{M}_a\,\dot{M}_a\,,
\end{eqnarray}
from which one recognizes on comparison with
(\ref{h00candots})-(\ref{Mijrad}) that
\begin{equation}\label{A10}
\eta=\frac{2918}{735}+\frac{116}{21}\ln\left(\frac{r}{r_0}\right)\,.
\end{equation}

\acknowledgments

L.B. and B.R.I. thank the Indo-French collaboration IFCPAR under which
this work has been carried out. L.B. was invited by the Yukawa
Institute for Theoretical Physics in Kyoto during the last phase of
this work.

\bibliography{BDI04zeta} \end{document}